\documentclass[usenatbib,iop,numberedappendix]{aeb_emulateapj_2010}
\usepackage{amsmath}
\usepackage{amssymb}
\usepackage{xspace}
\usepackage[normalem]{ulem}
\usepackage{mathrsfs}

\usepackage{natbib}

\usepackage[usenames,dvips]{color}

\def\del#1{{}}

\def\s{{\rm s}} %...........seconds
\def\pc{{\rm pc}} %.........parsecs
\def\Mpc{{\rm M}\pc} %......megaparsecs
\def\erg{{\rm erg}} %.......ergs
\def\lnls{$\log \mathcal{N}$-$\log S$\xspace}
\def\N{\mathcal{N}}
\def\F{\mathcal{F}}
\def\Fermi{{\em Fermi}\xspace}
\def\QLF{\phi_Q(z,L)}

\begin{document}

\title{
Implications of Plasma Beam Instabilities for the Statistics of the Fermi Hard Gamma-ray Blazars and the Origin of the Extragalactic Gamma-Ray Background
}

\author{
Avery E.~Broderick\altaffilmark{1,2},
Christoph Pfrommer\altaffilmark{3},
Ewald Puchwein\altaffilmark{3}, and 
Philip Chang\altaffilmark{4}
}
\altaffiltext{1}{Perimeter Institute for Theoretical Physics, 31 Caroline Street North, Waterloo, ON, N2L 2Y5, Canada}
\altaffiltext{2}{Department of Physics and Astronomy, University of Waterloo, 200 University Avenue West, Waterloo, ON, N2L 3G1, Canada}
\altaffiltext{3}{Heidelberg Institute for Theoretical Studies, Schloss-Wolfsbrunnenweg 35, D-69118 Heidelberg, Germany}
\altaffiltext{4}{Department of Physics, University of Wisconsin-Milwaukee, 1900 E. Kenwood Boulevard, Milwaukee, WI 53211, USA}

\shorttitle{A Unified Model of the Hard Gamma-ray Blazars}
\shortauthors{Broderick et al.}

\begin{abstract}
  \Fermi has been instrumental in constraining the luminosity function and
  redshift evolution of gamma-ray bright blazars.  This includes
  limits upon the spectrum and anisotropy of the extragalactic
  gamma-ray background (EGRB), 
  redshift distribution of nearby \Fermi active galactic nuclei (AGN), and the
  construction of a \lnls~relation.  Based upon these, it has been argued that
  the evolution of the gamma-ray bright blazar population must be much less
  dramatic than that of other AGN.  However, critical to such claims is the
  assumption that inverse Compton cascades reprocess emission above a TeV into
  the \Fermi energy range, substantially enhancing the strength of the observed
  limits.  Here we demonstrate that in the absence of such a process, due, e.g.,
  to the presence of virulent plasma beam instabilities that preempt the
  cascade, a population of TeV-bright blazars that evolve similarly to quasars
  {\em is} consistent with the population of hard gamma-ray blazars observed by
  \Fermi.  Specifically, we show that a simple model for the properties and
  luminosity function is simultaneously able to reproduce their \lnls~relation,
  local redshift distribution, and contribution to the EGRB and its
  anisotropy {\em without any free parameters}.  Insofar the
  naturalness of a picture in which the hard gamma-ray blazar
  population exhibits the strong redshift evolution observed in other
  tracers of the cosmological history of accretion onto halos is
  desirable, this lends support for the absence of the inverse Compton
  cascades and the existence of the beam plasma instabilities.
\end{abstract}

\keywords{BL Lacertae objects: general -- gamma rays: general -- radiation mechanisms: non-thermal}

\maketitle

\section{Introduction} \label{sec:I}

\subsection{Blazars in the \Fermi era}
The Large Area Telescope (LAT) onboard the \Fermi gamma-ray space
telescope has become a powerful tool for studying
gamma-ray bright active galactic nuclei (AGNs), placing the most
stringent constraints to date upon their numbers and evolution.  In
practice, this is performed via a variety of methods, including the
flux and redshift distributions of nearby sources, and the
extragalactic gamma-ray background (EGRB) due to unresolved sources at
high redshift.  Each of these effectively probes different projections
of the evolving luminosity function of the gamma-ray bright objects,
and thus taken together provides considerable traction upon their
populations at low and high redshifts.

The \Fermi~AGN sample is overwhelmingly dominated by blazars, with a
handful of radio and starburst galaxies comprising the remainder
\citep[see, e.g., Table 5 of ][]{2LAC}.  The population of blazars is
itself often sub-divided into a number of sub-categories, depending
primarily upon their optical properties.  The most populated are the
flat-spectrum radio sources (FSRQs) and BL Lacs, both of which are
further segregated into low, intermediate, and high synchrotron peak
sources (LSP, ISP, HSP, respectively).  The latter categories roughly
correspond to the hardness or softness of the gamma-ray spectrum, 
$E^2 dN/dE$, at energies relevant for \Fermi, with HSPs being harder
than ISPs which are harder than LSPs.
Typically, the FSRQs are considerably softer than the BL Lacs, and
thus appear primarily at low energies ($\lesssim 10$~GeV).  
In contrast, a number of BL Lacs exhibit rising spectra between 1 GeV
and 100 GeV, which we define as ``hard gamma-ray blazars''.

The rising \Fermi~spectra of the hard gamma-ray blazars suggest a
natural identification with the observed set of TeV blazars, detected
and characterized by imaging atmospheric Cerenkov telescopes such as
H.E.S.S., VERITAS, and MAGIC\footnote{High Energy Stereoscopic System,
  Very Energetic Radiation Imaging Telescope Array, Major Atmospheric
  Gamma Imaging Cerenkov Telescope.}.  As for \Fermi, the
extragalactic TeV universe is dominated by blazars\footnote{For an
  up-to-date list, see
  http://www.mppmu.mpg.de/$\sim$rwagner/sources.}: of the 28 objects
with well-defined spectral energy distributions (SEDs) listed in
\citet{PaperI}, 24 are blazars, which we refer to as the  
``TeV blazars''.  Thus, any 
limitation upon the evolution of the hard gamma-ray blazar population
implies a corresponding constraint upon the TeV blazars, and vice
versa. 

The TeV blazars are all relatively nearby, with $z\sim0.1$
typically.  This is the result of the annihilation of TeV gamma-rays
upon the extragalactic background light (EBL), and the subsequent
generation of a relativistic population of pairs
\citep{Goul-Schr:67,Sala-Stec:98,Nero-Semi:09}.  Typical gamma-ray
mean free paths range from 30 Mpc to 1 Gpc, depending upon gamma-ray
energy and source redshift, explaining the paucity of high-redshift
TeV sources.  The subsequent evolution of the energetic population of
pairs is subject to two competing scenarios.  

Historically, it has been assumed that
these cool primarily by Comptonizing the cosmic microwave background,
resulting in an inverse Compton cascade that effectively reprocesses
the original TeV emission to energies below 100 GeV.  The assumption
that this reprocessing occurs has a dramatic impact upon the
implications \Fermi~has for the TeV blazar population. 
In this first scenario, stringent constraints on the number of
hard gamma-ray blazars at high redshift can be derived from the
EGRB, for which \Fermi~has
provided the most precise estimate.  Even based upon EGRET data, it
has been well established that in the presence of inverse Compton
cascades the TeV blazar population cannot exhibit the dramatic
evolution that characterizes other AGN specifically, 
and other tracers of the cosmological history of accretion onto
galactic halos more generally (e.g., star formation), with the most 
recent \Fermi~EGRB limits implying that their co-moving number density
be essentially fixed 
\citep{Naru-Tota:06,Knei-Mann:08,Inou-Tota:09,Vent:10}.  This
represents a substantial obstacle to unifying the hard gamma-ray
blazar population with that of other AGN, is at
odds with the underlying physical picture of accreting black hole
systems, and suggests an unlikely conspiracy between accretion physics
and the formation of structure.

In a series of papers 
\citep[][hereafter Paper I, Paper II, Paper III]{PaperI,PaperII,PaperIII} 
and \citet{PaperIV}, we have explored the possible impact of
beam-plasma instabilities upon the gamma-ray emission of bright TeV 
sources and their subsequent cosmological consequences. 
We found that in this second scenario a
variety of cosmological puzzles, most importantly the statistics of
the high-redshift Ly$\alpha$ forest, were naturally resolved if the
VHEGR emission from TeV blazars was dumped into heat in the
intergalactic medium, as anticipated by such plasma
instabilities\footnote{In practice, the ability of plasma
  instabilities to efficiently thermalize the pairs' kinetic energy
  depends upon their nonlinear evolution.  This is presently highly
  uncertain \citep[see, e.g., Paper I,][]{Schl_etal:12,Mini-Elyv:12}.
  Here we assume only that the inverse Compton cascades are preempted,
  presumably by such plasma instabilities, and explore the
  consequences for the \Fermi~hard gamma-ray blazar population and EGRB.}.
However, to do so requires a much more rapidly rising TeV blazar
co-moving number density than implied by previous 
analyses of the EGRB, namely one similar to that of quasars and
qualitatively consistent with these other examples that depend on the
cosmological history of accretion (e.g., star formation, radio
galaxies, AGNs, galactic merger rates, etc.). 
As shown in Paper I, the apparent tension with the above discussed
constraints from the EGRB is reconciled by the lack of 
significant inverse Compton cascades, preempted by the plasma
instabilities responsible for depositing the VHEGR luminosity into the
intergalactic medium.  Without the inverse Compton cascades, it is
possible to quantitatively reproduce both the redshift-dependent
number of hard gamma-ray blazars listed in the First Fermi LAT
AGN Catalog \citep[1LAC, ][]{Fermi_AGNCatalogue2010}, and the EGRB
spectrum above 10 GeV.

The recent release of the 2 Year \Fermi-LAT AGN Catalog
\citep[2LAC,][]{2LAC} and the First \Fermi-LAT Catalog of $>10$~GeV
sources \citep[1FHL,][]{1FHL}, motivates a reevaluation of the
\Fermi~constraints upon the evolution of the number and luminosity
distribution of the hard gamma-ray blazars within the context of a
considerably more complete set of resolved \Fermi~sources.
With the luminosity function posited in Paper I, here we 
explicitly construct the expected flux and redshift distributions, and
the \Fermi~EGRB, and directly assess the viability of a quasar-like
evolution in the hard gamma-ray blazar population.  Generally, we find
excellent agreement where expected, implying that in the absence of
inverse Compton cascades (preempted, e.g., by plasma instabilities) it
is possible to unify the hard gamma-ray blazars with AGN generally.

\subsection{Methodology and Outline}

Our primary goal is to observationally probe the redshift-dependent
luminosity function of the TeV blazars.  In practice, this is
complicated by the small number and limited redshift range of the
known TeV blazars.  The sample of observed TeV blazars is
strongly biased in favor of nearby, X-ray selected BL Lac
objects. To address those selection effects, we use the
\Fermi~hard gamma-ray blazars (defined by an {\em intrinsic} photon
spectral index $\le2$) as proxies.

At low redshift ($z\lesssim0.2$), the cosmological redshift and
absorption on the EBL are negligible below 100 GeV, and the two
blazar populations are directly comparable.  We exploit this to
empirically define the distribution of intrinsic spectra relevant
for the hard gamma-ray blazars, extending the luminosity function
described in Paper I and removing a key degeneracy therein.

However, even at moderate redshifts ($z\gtrsim0.2$) the 
absorption on the EBL substantially softens the spectra below
100 GeV \citep{Fermi_EBL2012}, and this must be taken into account in the source
identification.  Where direct comparisons to the \Fermi~blazar sample
are made, we make the conservative choice of considering only objects
with {\em observed} photon spectral indexes $\le2$, which necessarily
implies that the {\em intrinsic} photon spectral indexes are also
$\le2$ (Sections \ref{sec:lNlS} and \ref{sec:dNdz}). For
concreteness, we define the ``hard \Fermi~blazars'' to be those objects 
that exhibit a rising spectrum, with index $\Gamma_F\le 2$ in the energy 
band 1--100 GeV, where $E^2 dN/dE\propto E^{2-\Gamma_F}$.
 For consistency it is necessary to restrict the
expected source population as well, and therefore we construct an
approximate relationship between the observed and intrinsic photon
spectral indexes.  Where comparison with the \Fermi~blazar sample is
not required, e.g., for modeling of the extragalactic gamma-ray
background, we consider the full hard gamma-ray population,
restricting only the intrinsic spectra (Section \ref{sec:EGRB}).

In Section \ref{sec:phiB} we define the TeV blazar luminosity function, describe
its regime of validity, and relate it to the luminosity function of
the \Fermi~hard gamma-ray blazars generally.  In
Section \ref{sec:comp} we review the definitions of the various
\Fermi~constraints and compare the expectations from our TeV blazar
luminosity functions.  Finally, discussion and conclusions are
contained in Section \ref{sec:C}.

\section{The Hard Gamma-ray Blazar Luminosity Function} \label{sec:phiB}

Here we construct a luminosity function for the hard gamma-ray
blazars, beginning with a review of the luminosity function for the
TeV blazars constructed in Paper I.  Critical to producing an
analogous luminosity function for the \Fermi~hard gamma-ray blazars is
the relationship between the \Fermi~band (here 100 MeV--100 GeV) and
the intrinsic isotropic equivalent TeV luminosity (100 GeV -- 10
TeV)\footnote{In practice, blazars are highly beamed, and thus the
  true intrinsic luminosity is reduced by the appropriate beaming
  factor.  However, this beaming factor is degenerate with the
  over-all normalization of the blazar number, with smaller beams
  offset by correspondingly larger intrinsic numbers.  Thus, in the
  interest of simplicity, here we consider only the isotropic
  equivalent luminosities.}.

\subsection{The TeV Blazar Luminosity Function}
The vast majority of extragalactic TeV sources have also been
identified by \Fermi, and thus there is a close relationship between
the TeV blazars and the \Fermi~hard gamma-ray blazars (defined
explicitly below). 

The TeV blazars typically have falling SEDs above a TeV, with the
brightest sources having a photon spectral index of $\Gamma_{\rm
  TeV}\simeq3$ (where $\Gamma_{\rm TeV}$ is defined by $dN/dE\propto
E^{-\Gamma_{\rm TeV}}$ from 100 GeV--10 TeV), implying a peak in the
SED at energies $\lesssim1$ TeV.  Below 100 GeV these sources are
among the hardest in the \Fermi~AGN sample, with rising SEDs, implying
a peak above 100 GeV.

In principle, we define the TeV-band luminosity function of TeV
blazars by 
\begin{equation}
\tilde{\phi}_B(z,L_{\rm TeV}) = \frac{d\N}{d\log_{10}L_{\rm TeV}\,d^3\!x}\,,
\end{equation}
where $\N$ is the number of TeV blazars with isotropic equivalent TeV luminosities above
$L_{\rm TeV}$, and in keeping with the notation in Papers I-III we
denote quantities defined in terms of physical volumes by tildes (as
opposed to co-moving volumes).

Measuring $\tilde{\phi}_B$ in practice is complicated by the large
optical depth to annihilation on the EBL for gamma rays with energies
above 100 GeV.  The pair-production mean free path is both energy and
redshift dependent, locally given by (Paper I)
\begin{equation}
D_{\rm pp}(E,z) = 35\left(\frac{E}{1~{\rm TeV}}\right)^{-1}\left(\frac{1+z}{2}\right)^{-\zeta}~{\rm Mpc}\,,
\end{equation}
where $\zeta=4.5$ for $z<1$ and $\zeta=0$ for $z\ge1$
\citep{Knei_etal:04,Nero-Semi:09}.  The redshift evolution is due to
the EBL, and is sensitive primarily to the star formation history.
The associated optical depth for a gamma-ray emitted at redshift $z$
and {\em observed} at an energy $E_{\rm obs}$, is then
\begin{equation}
\tau(E_{\rm obs},z)
\equiv
\int_0^z
\frac{dD_p}{dz'}
\frac{dz'}{D_{\rm pp}\left[ E_{\rm obs} (1+z'),z' \right]}\,,
\end{equation}
where $D_P\equiv \int c dt' = \int c dz/[H(z)(1+z)]$ is the proper
distance\footnote{In the definition of the Hubble function, we adopt the
  WMAP7 parameters, $H_0=70.4~{\rm km~s^{-1}~Mpc^{-1}}$,
  $\Omega_m=0.272$ and $\Omega_\Lambda=0.728$, in terms of which,
  $H(z)^2 = H_0^2\left[(1+z)^3\Omega_m + (1+z)^2(1-\Omega_m-\Omega_\Lambda)+\Omega_\Lambda\right]$.}.
At 1 TeV this is unity at a redshift of $z\simeq0.14$, and
TeV blazars are visible at only low redshifts, preventing a direct
measurement of the evolution of $\tilde{\phi}_B$.

The existing collection of TeV blazars is the result of targeted
observations, motivated by features in other wavebands, and is
therefore subject to a number of ill-defined selection effects.
Nevertheless, in Paper I we constructed an approximate luminosity
function for these objects at $z\sim0.1$.  It was found that this was
in excellent agreement with the quasar luminosity function, $\tilde{\phi}_Q$,
given by \citet{Hopkins+07}, and summarized in Appendix \ref{app:QLF},
upon rescaling the bolometric luminosity and overall normalization:
\begin{equation}
\tilde{\phi}_B(0.1,L_{\rm TeV})
\simeq
3.8\times10^{-3}\tilde{\phi}_Q(0.1,1.8L_{\rm TeV})\,.
\label{eq:BLFlocal}
\end{equation}

Included in this are a variety of uncertain corrections for various
selection effects.  Previously, we have attempted to estimate these by
identifying the TeV blazars with the \Fermi~hard gamma-ray blazars.
Within the context of the 2LAC we reconsider these, focusing upon the
duty cycle ($\eta_{\rm duty}$) and source selection ($\eta_{\rm sel}$)
corrections.  There remains considerable uncertainty in the relevant
source populations to compare, however.  The TeV blazars are
necessarily at very low redshift, suggesting that we should compare
them only to the nearby \Fermi~hard gamma-ray blazar population.  Less
clear is what redshift cut to impose.  At $z=0.1$, 0.15, and 0.2,
there are 9, 14, and 17 TeV blazars and 16, 37, and 49 \Fermi~hard
gamma-ray blazars with measured redshifts in the 2LAC,
respectively (i.e., above the catalog's flux limit).  Noting that
nearly all of the TeV blazars have now been detected by \Fermi, this
implies that the selection bias induced by the incomplete sky and time
coverage of TeV observations requires a correction factor of 1.8 to
2.9.  Furthermore, of the 277 \Fermi~hard gamma-ray blazars in the
2LAC, only 110 have measured redshifts.  Assuming these are drawn from
the same underlying population, this provides an additional correction
of $2.5$\footnote{This estimate should be taken with some caution,
  however.  In Paper I we found that based upon their spectral index
  and flux distributions, the population without redshifts were more
  consistent with being drawn from lower redshifts ($z<0.25$) than
  higher redshifts.  This would increase the normalization somewhat.}.
In combination, the associated selection correction ranges from 4.5 to
9.8.  In Paper I, and implicitly employed in Equation
(\ref{eq:BLFlocal}), we assumed $\eta_{\rm sel}\times\eta_{\rm
  duty}=6.4$, intermediate to those inferred from the above.  However,
the origin of this factor is rather different: the decrease in
$\eta_{\rm duty}$ to unity has been nearly exactly offset by the
increase in the overall number of \Fermi sources, and thus in
$\eta_{\rm sel}$.  Hence, the numerical factors in equation
(\ref{eq:BLFlocal}), as derived in Paper I, remain unchanged.

Motivated by the strong similarities with the local quasar luminosity
function, we posited that this relationship held at large $z$ as
well.  This has received indirect circumstantial support via the
observational consequences of the plasma-instability induced heating
of the intergalactic medium described in Papers II, III and
\citet{PaperIV}.  Of particular note is the great success in the
quantitative reproduction of the high-$z$ Ly$\alpha$ forest.

\subsection{Relationship to The \Fermi~Blazars}
While relating the TeV blazars and the \Fermi~hard gamma-ray blazars
is natural in principle, some care must be taken in practice.
Difficulties arise from the uncertain relationship between the
observed \Fermi-band fluxes and $L_{\rm TeV}$, the annihilation of
the high-energy gamma rays, and the distribution of source
properties.  Here we assume a specific family of SEDs for the TeV
blazars, use these to define the associated \Fermi~observables, and
discuss the inherent restrictions upon the \Fermi~blazar population
implied by these choices.

\subsubsection{Intrinsic \Fermi~Hard Gamma-ray Blazar SED} \label{sec:IS}

Relating the fluxes above a TeV and at energies relevant for
\Fermi~($\lesssim100$ GeV), requires some knowledge about the
intrinsic SED of the TeV blazars.  As already mentioned, the SED above
a TeV is slowly falling, with a photon spectral index of 3 typical
(i.e., $E^2 dN/dE\propto E^{-1}$).  However, for the two brightest TeV
blazars in the sky, Mkn 421 and 1ES 1959+650, the \Fermi~photon
spectral indexes, defined from 1 GeV--100 GeV, are $\Gamma_F=1.77$ and
1.94, respectively (i.e., $E^2 dN/dE\propto E^{0.2}$).  This spectral
shape is generic; TeV sources with spectra well characterized by
\Fermi~below 100~GeV show a median shift between their photon spectral
indexes below 100~GeV and above 1~TeV of 1.2
\citep[see Figure 44 of][]{2LAC} and the typical $\Gamma_F$ for the hard
gamma-ray blazars is $\sim1.8$ (see below).  Thus, it is clear
empirically that a single power law is a poor model for the intrinsic
SED \citep[see, e.g., Figures 13 \& 21 of][]{Fermi_SED2010}.

A more complicated SED is further motivated theoretically by the
identification of the high-energy gamma-ray emission with the
Comptonized synchrotron bump.  Models of the blazar spectrum exhibit a
peak near the TeV for the TeV-bright objects, suggesting that a
similarly peaked SED must be considered in practice.
Here, we model the intrinsic SED as a family of broken power laws:
\begin{equation}
\frac{dN}{dE} 
= 
f
\left[ 
  \left(\frac{E}{E_b}\right)^{\Gamma_l} 
  + 
  \left(\frac{E}{E_b}\right)^{\Gamma_h}
\right]^{-1}\,,
\label{eq:SED}
\end{equation}
for some normalization $f$ (with units of ${\rm photons~GeV^{-1}~s^{-1}}$, and
where we will set $E_b=1~{\rm TeV}$, $\Gamma_h\simeq\Gamma_{\rm TeV}=3$,
chosen as typical values.

\begin{figure}
\begin{center}
\includegraphics[width=\columnwidth]{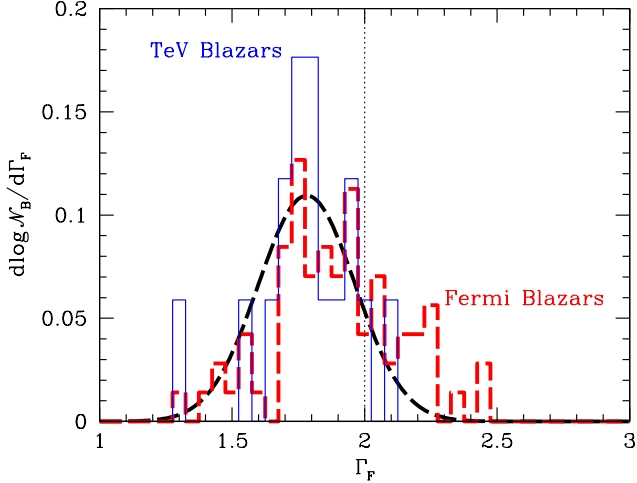}
\end{center}
\caption{Distribution of the \Fermi~photon spectral index for TeV
  blazars (blue solid) and all of the \Fermi~blazars (red dashed) with
  $z<0.2$.  The adopted intrinsic $\Gamma_l$ distribution, a
  Gaussian fit with mean $1.78$ and standard deviation $0.18$, is
  shown by a dashed line.  For reference, the spectral cut defining the
  hard gamma-ray blazars is shown by the vertical dotted
  line.\\}\label{fig:dNdGF}
\end{figure}

The choice of $\Gamma_l$ is complicated by the fact that some of the
observable tests described in Section \ref{sec:comp} are sensitive to
its value.  However, we have some observational guidance in
the form of the \Fermi~photon spectral indexes for the TeV blazars
themselves.  Figure \ref{fig:dNdGF} shows the $\Gamma_F$ distribution
of the nearby TeV blazars and the \Fermi~hard gamma-ray blazars.
The former is well fit by a Gaussian, with mean
$\bar{\Gamma}_l=1.78$ and standard deviate $\sigma_l=0.18$.  Based 
upon this we adopt the expanded luminosity function:
\begin{equation}
\begin{aligned}
\tilde{\phi}_B(z,L_{\rm TeV},\Gamma_l)
&\equiv
\frac{d\N}{d\log_{10}L_{\rm TeV}\,d^3\!x\,d\Gamma_l}\\
&=
\tilde{\phi}_B(z,L_{\rm TeV}) \frac{e^{-(\Gamma_l-\bar{\Gamma}_l)^2/2\sigma_l^2}}{\sqrt{2\pi}\sigma_l}\,.
\end{aligned}
\label{eq:eBLF}
\end{equation}
Note that the $\Gamma_F$ distribution of the TeV blazars is somewhat
harder than that of the hard gamma-ray blazars.
This is likely due to the dramatic drop in TeV luminosity when the
location of the Compton peak falls well below 100 GeV.
We discuss this point, and the limitation it implies, in
more detail in Section \ref{sec:limits}.

\subsubsection{Relating the TeV blazars and the hard gamma-ray blazars} \label{sec:rTF}
The luminosity function in Equation (\ref{eq:eBLF}) is still presented
in terms of intrinsic quantities (e.g., $\Gamma_l$, $L_{\rm TeV}$,
etc.).  However, frequently it will be necessary to relate these to
quantities that are directly measurable by \Fermi.  These will be
impacted both by the redshifting of the intrinsic spectrum and the
gamma-ray annihilation on the EBL.

\begin{figure}
\begin{center}
\includegraphics[width=\columnwidth]{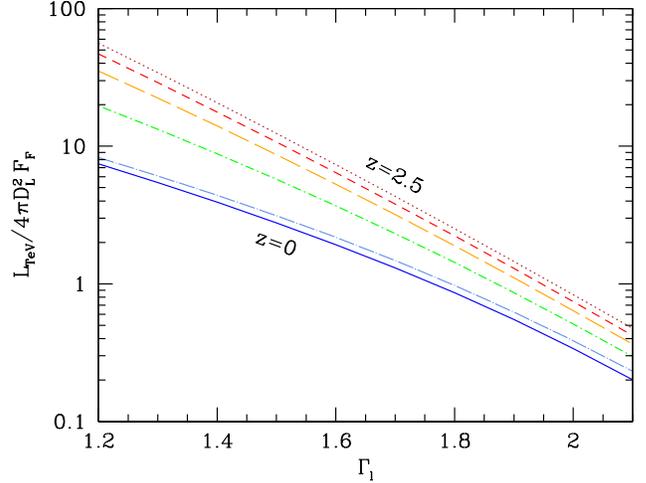}
\end{center}
\caption{Ratio of the intrinsic TeV-band luminosity and the observed
  \Fermi-band luminosity (100 MeV--100 GeV) as a function of
  $\Gamma_l$ for various redshifts, ranging from $0$ to $2.5$ in steps
  of $0.5$.}\label{fig:LFf}
\end{figure}

It is straightforward to show that the flux and fluence observed by \Fermi~between
energies $E_m$ and $E_M$ (e.g., 100 MeV and 100 GeV)
from a source at redshift $z$ is
\begin{equation}
F_F
=
\frac{1}{4\pi D_L^2}
\int^{(1+z)E_M}_{(1+z)E_m} dE \,E \frac{dN}{dE}\, e^{-\tau[E/(1+z),z]}\,,
\label{eq:FF}
\end{equation}
and
\begin{equation}
\F_F
=
\frac{1+z}{4\pi D_L^2}
\int^{(1+z)E_M}_{(1+z)E_m} dE \,\frac{dN}{dE}\, e^{-\tau[E/(1+z),z]}\,,
\label{eq:fF}
\end{equation}
respectively, where $D_L$ is the luminosity distance.
Similarly, the intrinsic TeV luminosity is, 
\begin{equation}
L_{\rm TeV}
=
\int_{0.1~{\rm TeV}}^{10~{\rm TeV}} dE \,E\frac{dN}{dE}\,.
\end{equation}
Thus, we have a redshift and SED-dependent relationship between $F_F$ and
$L_{\rm TeV}$:
\begin{equation}
\frac{L_{\rm TeV}}{4\pi D_L^2 F_F}(z,\Gamma_l)
=
\frac{
  \int_{0.1~{\rm TeV}}^{10~{\rm TeV}} dE \,E \,(dN/dE)
}{
  \int^{(1+z)E_M}_{(1+z)E_m} dE \,E \,(dN/dE)\,e^{-\tau[E/(1+z),z]}
}\,,
\label{eq:LTeV}
\end{equation}
where the denominator is simply the isotropic equivalent Fermi-band
luminosity.  This is shown for a handful of redshifts as a function of
$\Gamma_l$ in Figure \ref{fig:LFf}.  Note that this neglects any
inverse Compton cascade component, which would otherwise increase
$F_F$ beyond the intrinsic emission.

\begin{figure}
\begin{center}
\includegraphics[width=\columnwidth]{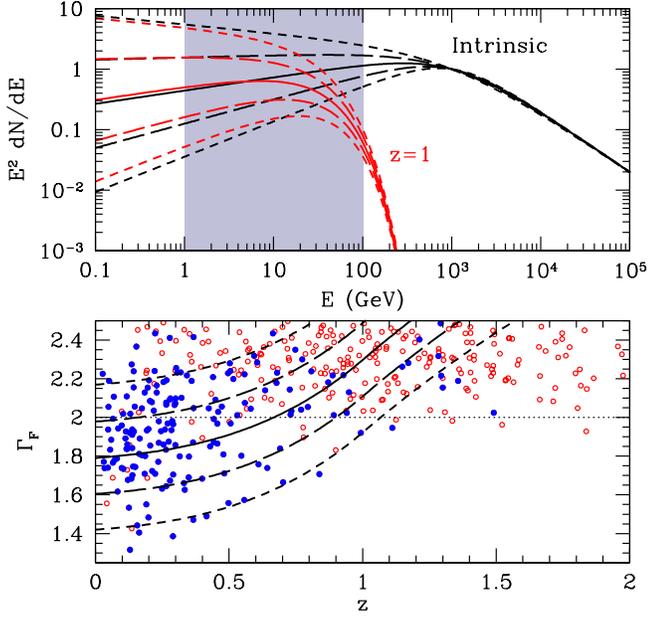}
\end{center}
\caption{Top: Example intrinsic (black) and observed (red) SEDs at
  $z=1$ for $\Gamma_l=\bar{\Gamma}_l$ (solid),
  $\Gamma_l=\bar{\Gamma}_l\pm\sigma_l$ (long dash), and
  $\Gamma_l=\bar{\Gamma}_l\pm2\sigma_l$ (short dash).  The
  normalization is arbitrary, and set here so that $E^2 dN/dE=1$ at
  the break energy (1 TeV).  For reference the energy range over which
  $\Gamma_F$ is defined is shaded.  Bottom: Inferred $\Gamma_F$ for
  the spectra in the top panel, with corresponding line types.  For
  reference, the photon spectral index cutoff that defines the hard
  gamma-ray blazars is shown by the dotted line.  The photon spectral
  indexes of the \Fermi~blazars are shown as a function of redshift by
  the points, with BL Lacs and non-BL Lacs (predominantly FSRQs)
  shown by the blue filled and red open circles.}\label{fig:GFz}
\end{figure}

Similarly important for the definition of the \Fermi~sources is the
\Fermi-band photon spectral index, $\Gamma_F$.  Again this is modified
by the redshift (different portions of the intrinsic spectrum are
being observed) and by absorption on the EBL.  Assessing the impact
these have upon the measured $\Gamma_F$ depends on how it
is defined.  Here we estimate $\Gamma_F$ via a least-squares
fit to the redshifted and absorbed intrinsic spectrum\footnote{In
  practice this is done via a linear fit in $\log dN/dE$ versus $\log
  E$.} between 1 GeV
and 100 GeV, the energy range over which it is defined in the 2LAC.
The impact upon $\Gamma_F$ is shown in Figure \ref{fig:GFz}.  At
high redshift even intrinsically hard spectra appear soft due to
absorption.  For example, a source at $z=0.667$ with
$\Gamma_l=\bar{\Gamma}_l=1.78$, will have a $\Gamma_F\simeq2$.  Thus, even
moderate redshifts are sufficient to move objects out of the hard
gamma-ray blazar class.

Also shown in Figure \ref{fig:GFz} is the evolution of the
distribution of $\Gamma_F$ for the \Fermi~blazars.  This may occur for
a variety of reasons, including a correlation between $\Gamma_F$ and
bolometric luminosity \citep[see, e.g., ][]{Ghis:11}.  Recently, it
has been shown explicitly that this cannot account for the entirety of
the spectral evolution, with absorption necessarily playing a role
\citep{Fermi_EBL2012}.  Here, we note simply that there is excellent
agreement between the lower envelope of the Fermi sources and the
evolution of the $\Gamma_F$ associated with the 2$\sigma$ lower limit
upon $\Gamma_l$ from the TeV blazars alone (shown by the short-dash
line).

\subsubsection{Limitations upon the Hard Gamma-ray Blazar Luminosity Function} \label{sec:limits}

Our empirical TeV blazar luminosity function necessarily was
constructed only for TeV-bright objects, and thus only
describes the TeV-bright blazar population.  As a consequence, some
care must be taken in extending this to the entire \Fermi~blazar population.
In particular, the TeV blazar luminosity function poorly constrains
the population of soft gamma-ray blazars.
To address this, here we restrict ourselves to the class of
\Fermi~blazars with flat or rising spectra, and thus to the objects
with intrinsic gamma-ray spectra likely to peak well above 100 GeV.  That is,
we consider only objects for which $\Gamma_l\le2$, for which the
intrinsic SED in Equation (\ref{eq:SED}) peaks around $E_b=1~{\rm TeV}$.
Empirically, this is evident from the lack of TeV blazars with
$\Gamma_F>2.1$.

Due to the spectral softening arising from absorption on the EBL and
redshift, this condition upon the intrinsic SED does not translate
into a unique condition upon the observed SED.  That is,
$\Gamma_l\le2$ does not generally imply that $\Gamma_F\le2$.  The
converse is, however, true: $\Gamma_F\le2$ does imply $\Gamma_l\le2$ 
generally, as may be seen immediately in Figure \ref{fig:GFz}.  Thus,
where we wish to construct populations of TeV-bright 
objects from the \Fermi~blazar sample for comparison with the hard
gamma-ray blazar luminosity function described in the previous section,
we will consider only the \Fermi~hard gamma-ray blazars.  This
includes the \lnls~relation and redshift distributions described in
Sections \ref{sec:lNlS} and \ref{sec:dNdz}, respectively.  

When we consider the TeV blazar contribution to the \Fermi~EGRB, we
do not require a corresponding population of \Fermi~blazars, and thus
retain only the more conservative condition upon $\Gamma_l$.  Concerns
regarding the generality of the associated high-energy EGRB are
discussed in Section \ref{sec:EGRB}, here we simply note that the
neglected population of soft gamma-ray blazars is significant only below a
few GeV.

\section{Comparisons with the \Fermi~Hard Gamma-ray Blazars} \label{sec:comp}
We now turn our attention to comparing the implications of the
luminosity function obtained in the previous section with various
measures of the \Fermi~hard gamma-ray population.  We note that the
specific properties of the luminosity function and the intrinsic
spectra are now completely defined, and thus in the following
comparisons to the \Fermi~blazar sample there are no degrees of
freedom to adjust.

Both the \lnls~and redshift distribution probe the recent evolution of
the hard gamma-ray blazar luminosity function.  Since they are both
one-dimensional, they are both necessarily projections of
$\tilde{\phi}_B$.  They differ in the form of the projection,
measuring in different degrees the redshift evolution and the
luminosity distribution of the hard gamma-ray blazars.  In contrast,
the \Fermi~isotropic EGRB is most sensitive to the unresolved
sources at high redshifts, and thus probes the peak of the luminosity
function in both redshift and luminosity.  While all are important,
the isotropic EGRB is likely to provide the most significant
constraint upon the viability of rapidly evolving hard gamma-ray
blazar luminosity functions.

\subsection{2LAC \lnls~Relation} \label{sec:lNlS}

The \lnls~relation describes the flux distribution of a particular
source class.  In it, $\N(S)$ is simply the number of sources with
fluxes $>S$, making it straightforward to define empirically.
Complications arise in selecting the particular source class of
interest, the definition of ``flux'' to be employed, and the treatment
of observational selection effects.  All of these are relevant for
\Fermi, and thus here we describe how we constructed the
\Fermi~\lnls~relation for the hard gamma-ray blazars and its relation
to the hard gamma-ray blazar luminosity function discussed in the
previous section.

\subsubsection{Observational Definition}

Depending upon application, the fluence from 100 MeV--100 GeV
($\F_{25}$), fluence from 1 GeV--100 GeV ($\F_{35}$), and flux from
100 MeV--100 GeV ($F_{25}$) have all been used as ``flux'' measures
for Fermi sources.  Primarily, $\F_{25}$ and $F_{25}$ have been used
to assess statistical properties of \Fermi~sources 
\citep[see, e.g.,][]{Fermi_AGNCatalogue2010,2LAC}.
This includes an empirical reconstruction of the \lnls~relation for the
\Fermi~blazars in the 1LAC in terms of $\F_{25}$ by
\citet{SPA12}.  However, within the 2LAC itself, $\F_{35}$, is the
flux measure reported.

We relate these here by assuming the spectrum across the Fermi LAT
band (100 MeV to 100 GeV) is well approximated by a single power law,
$dN/dE\,dt = f_F E^{-\Gamma_F}$, and therefore the fluence and flux are,
\begin{equation}
\F = f_F \int_{E_m}^{E_M} dE E^{-\Gamma_F}
=
\frac{f_F\left( E_m^{1-\Gamma_F} - E_M^{1-\Gamma_F} \right)}{\Gamma_F-1} \,,
\label{eq:fFdef}
\end{equation}
and
\begin{equation}
F =  f_F \int_{E_m}^{E_M} dE E^{1-\Gamma_F}
=
\frac{f_F\left( E_m^{2-\Gamma_F} - E_M^{2-\Gamma_F} \right)}{\Gamma_F-2} \,,
\label{eq:Fdef}
\end{equation}
respectively.\footnote{Here, we assumed $\Gamma_F \ne 1$ and $\Gamma_F \ne
  2$, respectively. While the first condition is empirically true, in
  case of $\Gamma_F=2$, we have $F=f_F \log(E_M/E_m)$.} The
normalization, $f_F$, is set by the reported $\F_{35}$, from which
$\F_{25}$ and $F_{25}$ may then be readily computed (see the Appendix
\ref{app:fdefs} for explicit expressions).

\begin{figure}
\begin{center}
\includegraphics[width=\columnwidth]{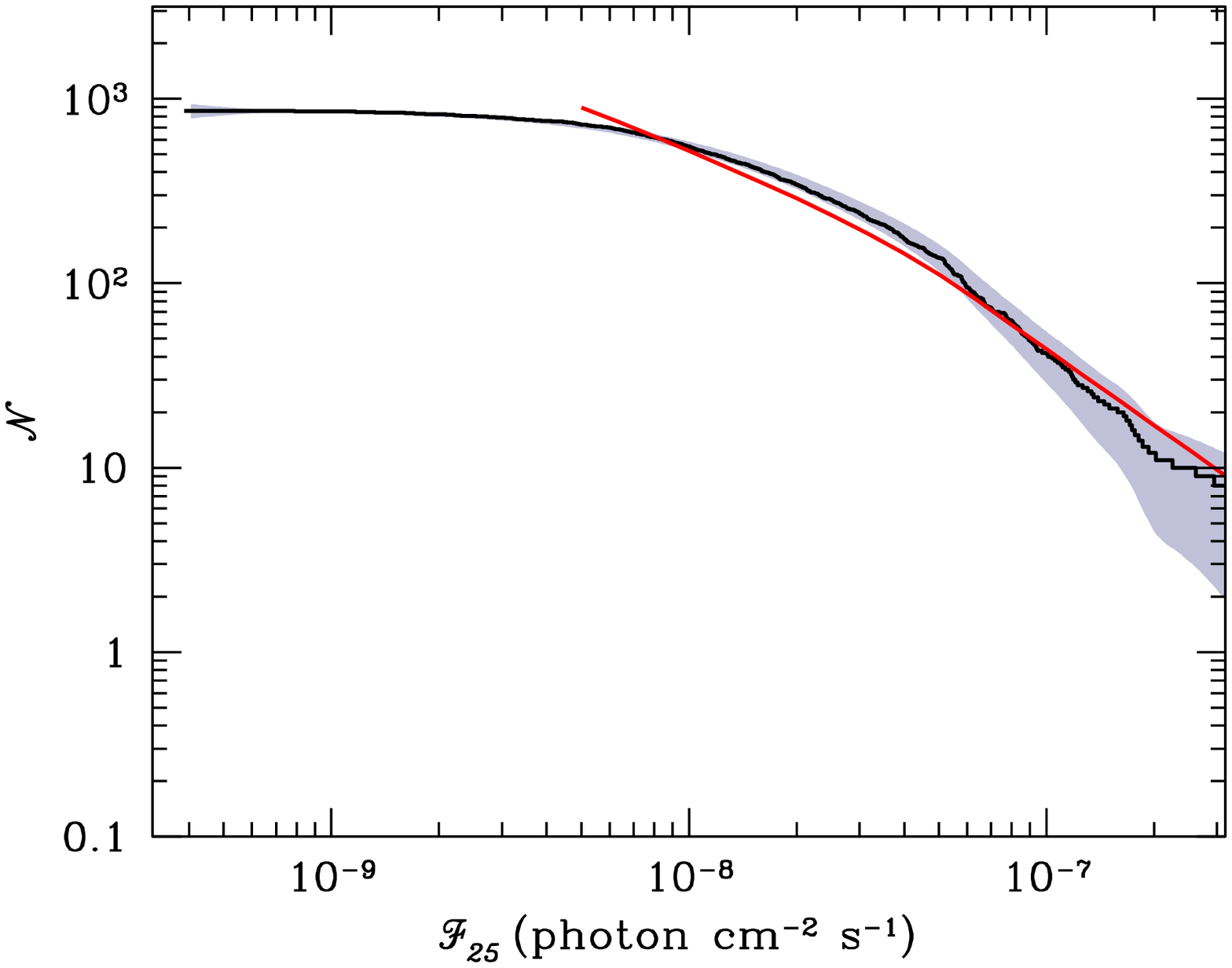}\\
\includegraphics[width=\columnwidth]{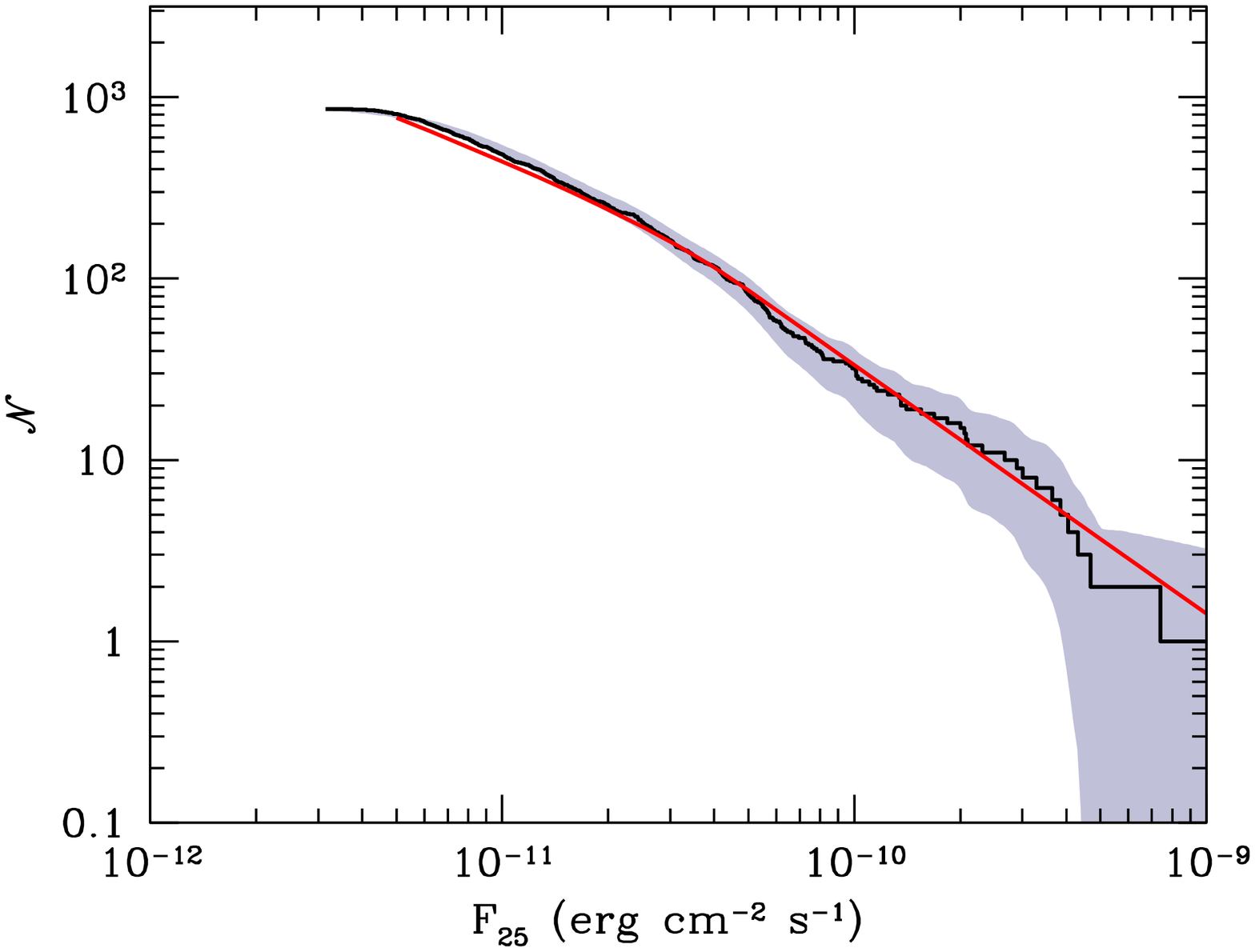}
\end{center}
\caption{\lnls~relation defined in terms of $\F_{25}$ (top) and
  $F_{25}$ (bottom) for the \Fermi~blazars in the 2LAC.  The shaded
  region provides an estimate of the 2$\sigma$ uncertainty in the
  \Fermi~\lnls~relation due to the measurement uncertainty on $F_{25}$
  and the Poisson fluctuations in the sample itself.  The
  empirically reconstructed \lnls~relation from \citet{SPA12} is shown
  by the red line.  For the latter, the \citet{SPA12} \lnls~relation
  has been inferred using the average spectral index, $2.13$.}\label{fig:lNlS_All}
\end{figure}

Figure \ref{fig:lNlS_All} shows the \lnls~relation for all of the
\Fermi~blazars in the 2LAC with a SNR $\ge7$, defined in terms of
$\F_{25}$ and $F_{25}$. 
The precision with which the \lnls~relation can be reconstructed
empirically is limited by both the intrinsic measurement uncertainty (in
$\F_{35}$ and $\Gamma_F$) and the limited number of AGN.  We attempt to
assess this uncertainty via a Monte Carlo simulation of the Fermi
catalog, using the reported measurement uncertainties (assuming a
normal and log-normal error distributions for $\Gamma$ and $\F_{35}$,
respectively) and constructing bootstrap samples of the 2LAC.
The 2$\sigma$ regions are shown by the gray shaded regions in Figures
\ref{fig:lNlS_All} and \ref{fig:lNlS}.  However, we note
that the errors at various fluxes are strongly correlated due to the
cumulative definition of $\N(S)$, and thus must be interpreted
cautiously.

The $F_{25}$ \lnls~relation is in good agreement with the
empirically constructed \lnls~relation from \citet{SPA12}, providing
some confidence in our reconstructed $F_{25}$ itself.  
Clearly evident in both forms of the \lnls~relation is a flattening at
small fluxes.  \citet{SPA12} identify this with a systematic 
bias induced by a correlation between the $\F_{25}$ flux limit and the
source spectral index, resulting in fewer soft sources being detected
below a few$\times10^{-8}\,{\rm photons~cm^{-2}~s^{-1}}$
\citep[see Figure 14 of ][]{2LAC}.  This results in a break in the
\lnls~relation roughly at the value for $\F_{25}$ at which the sample
becomes incomplete.  
Unlike $\F_{25}$, the flux limit in $F_{25}$ is only weakly dependent
upon $\Gamma_F$ \citep[cf. Figures 14 \& 15 in ][]{2LAC}, and the
break is correspondingly weaker. 

Despite the known bias, \citet{SPA12} have argued based upon the 1LAC
that the intrinsic \lnls~relation does indeed have a break near 
$\F_{25}=6\times10^{-8}~{\rm photons~cm^{-2}~s^{-1}}$.  We believe
this is suspect for three reasons.  First, its location is very near the
bias-induced break in the 1LAC.  Second, the location of the break in
the \lnls~relation constructed from the 2LAC blazars appears to have
moved towards marginally lower fluxes.  Third, the break is
considerably less prominent when a less biased flux is employed,
namely $F_{25}$.  This points to an as yet unidentified source of bias
for $F_{25}\lesssim 10^{-11}~{\rm erg~cm^{-2}~s^{-1}}$, and thus we
will restrict ourselves to fluxes above this cutoff.

The $F_{25}$ \lnls~relation for the \Fermi~hard gamma-ray blazars
specifically is shown in Figure \ref{fig:lNlS}.  Aside from the
restriction to blazars with 
$\Gamma_F<2$, this is constructed in an identical fashion to those
described above.  Apart from the reduced number of sources, it shares
many of the qualitative features found for the \lnls~relation from the
full blazar sample.  In particular, the same suspect flattening for
$F_{25}\lesssim 10^{-11}~{\rm erg~cm^{-2}~s^{-1}}$ is observed.

Associated with $\N(F_{25})$ is an estimate for the blazar
contribution to the unresolved Fermi background, arising from the
faint end of the blazar population:
\begin{equation}
F_{25,\rm tot}
=
\int_0^\infty d F_{25} \frac{d\N}{dF_{25}} F_{25} \,.
\end{equation}
This provides an independent constraint upon the overall
normalization, once resolved point sources have been removed.
The associated \Fermi~limit upon $\F_{25,\rm tot}$ is 
$18\pm2.4\times10^{-5}~{\rm ph~cm^{-2}~s^{-1}}$, while that
implied by the \lnls~ relation in \citet{SPA12} is
$11\times10^{-5}~{\rm ph~cm^{-2}~s^{-1}}$.  Note the condition
that $F_{25,\rm tot}$ be finite implies that $\N$ cannot be well
approximated by a single power law.  Above 
$F_{25}\simeq10^{-11}~{\rm erg~cm^{-2}~s^{-1}}$, \citet{SPA12} found
$\N\propto F_{25}^{-1.37\pm0.13}$, which were it to continue indefinitely to
small fluxes would imply that $F_{25,\rm tot}$ diverges at the faint
end.

\subsubsection{Relationship to $\tilde{\phi}_B$}
\label{sec:phi}

\begin{figure}
\begin{center}
\includegraphics[width=\columnwidth]{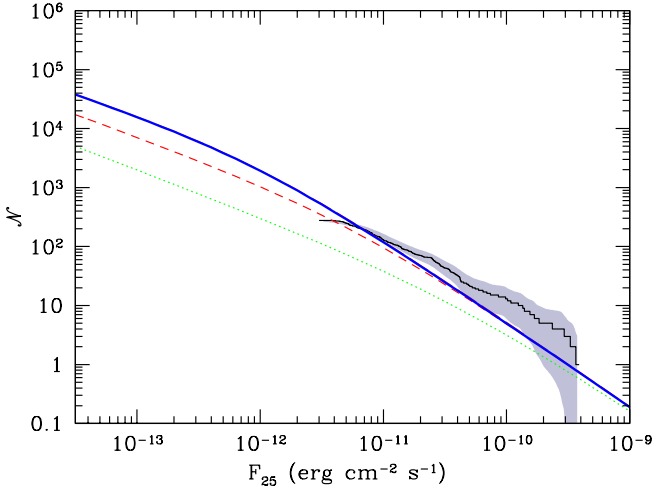}
\end{center}
\caption{\lnls~relation in terms of the flux from 100 MeV -- 100 GeV
  associated with the expanded hard gamma-ray blazar luminosity
  function presented in Equation (\ref{eq:eBLF}) in comparison with
  that from the 2LAC \Fermi~hard gamma-ray blazar sample.  The shaded
  region provides an estimate of the 2$\sigma$ uncertainty in the
  \Fermi~\lnls~relation due to the measurement uncertainty on $F_{25}$
  and the Poisson fluctuations in the sample itself.  The flattening
  in the \Fermi~\lnls~relation at low fluxes is probably an artifact
  of the \Fermi~flux limit.  For reference, the contributions to the
  \lnls~relation from hard gamma-ray blazars with $z\le0.1$ (green
  dotted) and  $z\le0.3$ (red dashed) are shown.}\label{fig:lNlS}
\end{figure}

The primary difficulty in producing a \lnls~relation to compare with
that constructed using the \Fermi~2LAC blazars is the treatment of the
particular selection effects relevant for the population of interest.
Specifically, it is necessary to produce cuts on $\Gamma_F$ and
$F_{25}$.  Thus, we define
\begin{multline}
\N
=
\eta_F
\int_0^2 d\Gamma_l
\int_0^\infty dz
\int_{\log_{10} L_{\rm TeV}(F_{25},z,\Gamma_l)}^\infty d\log_{10}L_{\rm TeV}\\
4\pi D_A^2\frac{dD_P}{dz} 
\Theta\left[2-\Gamma_F(\Gamma_l,z)\right]
\tilde{\phi}_B(z,L_{\rm TeV},\Gamma_l)\,,
\label{eq:lNlS}
\end{multline}
where $L_{\rm TeV}(F_{25},z,\Gamma_l)$ is given by Equation
(\ref{eq:LTeV}), $\Gamma_F(\Gamma_l,z)$ is obtained as described in
Section \ref{sec:rTF}, $\Theta(x)$ is the Heaviside function
(vanishing for $x<0$ and unity otherwise), and the cuts on $\Gamma_l$
and $\Gamma_F$ are motivated by Section \ref{sec:limits} (note that
the cut on $\Gamma_l$ is redundant).  The coefficient $\eta_F=0.826$
is the correction due to the sky-coverage of the \Fermi~clean sample
($|b|>10^\circ$, where $b$ is the Galactic latitude).
This is compared to the observed \Fermi~\lnls~relation for the hard
gamma-ray blazars in Figure \ref{fig:lNlS}.  

The cutoff in $\Gamma_F$ results in a $\Gamma_l$-dependent redshift
cut, which is exhibited as a break in the \lnls~relation that moves to
progressively larger fluxes as $\Gamma_l$ increases.  This is evident
in the \lnls~relations for sources restricted to smaller redshifts in
Figure \ref{fig:lNlS} (the green dotted and red dashed lines).  As
seen in Figure \ref{fig:GFz}, the location of this redshift limit is
sensitively dependent upon $\Gamma_l$, ranging between 0 and $\sim1$ for
$\Gamma_l$ from 2.0 to $\sim1.4$.  The $\Gamma_l$ sensitivity
of the location of this break is part of the
justification for the using the extended luminosity function in
Equation (\ref{eq:eBLF}).  
The location of the resulting break after
integrating over the TeV blazar $\Gamma_l$ distribution is near
$F_{25}=1.6\times10^{-12}~{\rm erg~cm^{-2}~s^{-1}}$, roughly a factor
of three below \Fermi's stated flux limit for the 2LAC, and a factor
of six below the point at which unknown systematic effects appear to
produce an artificial flattening of the \Fermi~\lnls~relation.

Above and below the break we obtain $\N\propto F_{25}^{-1.42}$ and
$\N\propto F_{25}^{-0.75}$, respectively.  Notably, despite the
difference in the location of the cutoff, both of the power
laws are consistent with those reported in \citet{SPA12}\footnote{Note
  that in \citet{SPA12}, the power law indexes are for $d\N/d\F_{25}$.}.
In the case of the latter, however, we suspect the agreement is
incidental.

Above a flux of $F_{25}=10^{-11}~{\rm erg~cm^{-2}~s^{-1}}$, the minimum
flux at which we trust the \Fermi~\lnls~relation, Equation (\ref{eq:lNlS})
reproduces the observed relation quite well.  This is especially true
for $F_{25}\lesssim10^{-10}~{\rm erg~cm^{-2}~s^{-1}}$.  At higher fluxes the
paucity of sources induces large Poisson errors, and thus the excess
bump at and above this flux is not significant.

Since we treat the hard gamma-ray blazar contributions to the EGRB in
detail in Section \ref{sec:EGRB}, here we simply note that the
anticipated contribution to the Fermi EGRB is
$\F_{25,\rm tot}\simeq 1.19\times10^{-5}~{\rm ph~cm^{-2}~s^{-1}}$.
This corresponds to roughly 6.6\% of the total \Fermi~EGRB from 100 
MeV to 100 GeV, and 11\% of that implied by the empirical
reconstruction from the 1LAC by \citet{SPA12}.  That the hard gamma-ray
blazars are responsible for a such a small fraction of the EGRB is not
surprising; below 10 GeV the EGRB is dominated by the FSRQs.
Nevertheless, even at 100 MeV we expect the hard gamma-ray blazars to
account for roughly 10\% of the EGRB.

\begin{figure}[t!]
\begin{center}
\includegraphics[width=\columnwidth]{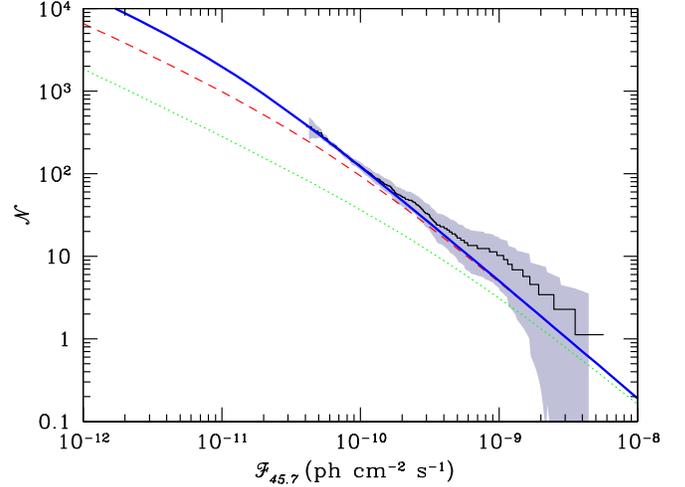}
\end{center}
\caption{\lnls~relation in terms of the fluence from 10~GeV--500~GeV
  ($\F_{45.7}$) associated with the expanded hard gamma-ray blazar
  luminosity function presented in Equation (\ref{eq:eBLF}) in
  comparison with 
  that from the 1FHL \Fermi~hard gamma-ray BL Lac
  sample.  The shaded region provides an estimate of the 2$\sigma$ uncertainty in the
  \Fermi~\lnls~relation due to the measurement uncertainty on $\F_{45.7}$
  and the Poisson fluctuations in the sample itself.  We have
  corrected for the detection efficiency following \citet{1FHL}, as
  described in Appendix \ref{sec:1FHLdeteff}, and as a consequence the
  \lnls~relation does not show the artificial flattening at low fluxes
  exhibited in Figures \ref{fig:lNlS_All} and \ref{fig:lNlS}.  For reference, the
  contributions to the \lnls~relation from hard gamma-ray blazars with
  $z\le0.1$ (green dotted) and  $z\le0.3$ (red dashed) are shown.}\label{fig:lNlS_1FHL}
\end{figure}

\subsection{1FHL \lnls~Relation} \label{sec:lNlS_1FHL}
Because the hard sources necessarily dominate at high energies, the
recently published 1FHL, a catalog of \Fermi~sources detected above
10~GeV, provides a means to probe the hard-source population
directly.  Already it is clear that for hard sources the flattening at
low fluxes is almost entirely, if not entirely, an artifact of the
LAT detection efficiency near the flux threshold 
\citep[see Appendix \ref{sec:1FHLdeteff} and Figures 31-33 of][]{1FHL}.  
Thus, there is currently no evidence for a break in the \lnls~relation
for the high-energy \Fermi~population.

As with the 2LAC sources, we may compare the \lnls~relation of 1FHL
sources with that anticipated by the expanded luminosity function in
Equation (\ref{eq:eBLF}).  Because the 1FHL reports the fluence
between 10~GeV and 500~GeV explicitly ($\F_{45.7}$), eliminating the
need to perform a spectral correction to this energy band, constructing
the observed \lnls~relation within this energy band is somewhat
simplified.  It is, however, complicated by the fact that the 1FHL
also includes a Galactic component that must be removed.  We do this
by considering only high latitude ($|b|>20^\circ$) sources that are
identified as BL Lac objects\footnote{These comprise roughly half of
  the 1FHL sample and are the dominant extragalactic component.}.
The resulting \lnls~relation is shown in Figure \ref{fig:lNlS_1FHL},
and is comparable to Figure 33 of \citet{1FHL}.

The anticipated \lnls~relation is constructed in a manner similar to
that in the previous section, replacing the relevant flux measure
with $\F_{45.7}$, and adjusting the correction to account for the
differing sky-coverage of the high-latitude 1FHL sample adopted.  In
addition, since the 1FHL detection efficiency is provided in
\citet{1FHL}, we make an effort to correct the \lnls~relation near the
detection threshold, as described in Section \ref{sec:1FHLdeteff}.
This is compared to the measured high-energy BL Lac \lnls~relation in
Figure \ref{fig:lNlS_1FHL}, providing an excellent fit over more than
an order of magnitude in fluence.
As with the 2LAC \lnls~relation shown in Figure \ref{fig:lNlS}, there
is an excess of sources at high fluxes in the 1FHL, though again this
is not significant.  We do predict a weak break near 
$\F_{45.7}\simeq2$--$3\times10^{-11}~{\rm ph~cm^{-2}~s^{-1}}$,
or roughly 50\% of the fluence of the dimmest 1FHL source, and hence
potentially accessible in the future.

\begin{figure}[t!]
\begin{center}
\includegraphics[width=\columnwidth]{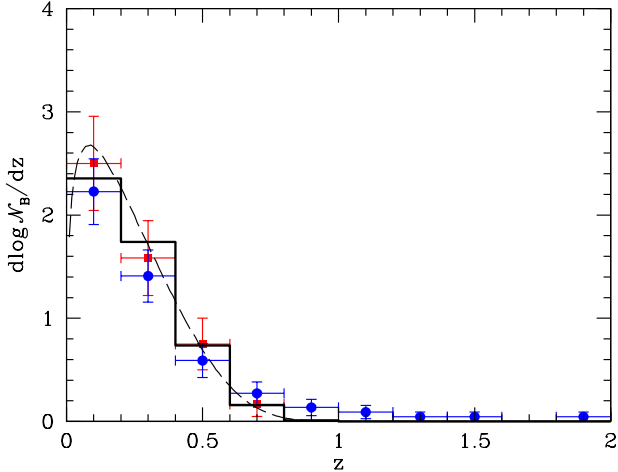}
\end{center}
\caption{Nearby redshift distribution of the hard gamma-ray
  blazars above the \Fermi~flux limit anticipated by the luminosity
  function in Equation (\ref{eq:eBLF}), both in continuous form (dashed)
  and binned with $\Delta z=0.2$ (continuous).  For comparison the
  redshift distribution of the \Fermi~hard gamma-ray blazars in the
  1LAC (red squares) and 2LAC (blue circles) are also shown.  For
  these, the vertical error bars denote Poisson errors and the
  horizontal error bars denote bin sizes.}\label{fig:dNdz} 
\end{figure}

\subsection{Hard Gamma-ray Blazar Redshift Distribution} \label{sec:dNdz}

In principle, the evolution in the number density of the nearby
\Fermi~hard gamma-ray blazars is directly probed by their observed
redshift evolution.  As with the \lnls~relation, this is
straightforward to define observationally.  However, in practice, it
is complicated by the flux-limited nature of the 2LAC, significantly
impacting even moderate redshifts, and the limited number of sources
with known redshifts (roughly 39\%).  Nevertheless, it represents a
different projection of the hard gamma-ray blazar luminosity function,
and provides a powerful additional test of the viability of a rapidly
evolving blazar population.

Within the context of the 1LAC, we demonstrated in Paper I that the
relatively large flux limit was capable of generating a precipitously
declining observed number of blazars, $\N_B$,
with redshift.  However, there we made a number of assumptions and
approximations regarding the intrinsic hard gamma-ray blazar spectra
and their relationship to $L_{\rm TeV}$.  Here we revisit this within
the more complete 2LAC and in terms of the more fully self-consistent
TeV blazar model described in Section \ref{sec:phiB}.  Particular
improvements over the computation in Paper I are the self-consistent
relationship between $L_{\rm TeV}$ and $F_{25}$, the distribution of
$\Gamma_l$, and the ability to now dispense with the upper limit upon
the TeV luminosity, to which our results are insensitive.

The definition of $\N_B(z)$ differs from $\N$ only by the limits of
integration:
\begin{multline}
\N_B(z)
=
\int_0^2 d\Gamma_l
\int_0^z dz'
\int_{\log_{10} L_{\rm TeV}(F_{25,\rm min},z',\Gamma_l)}^\infty d\log_{10}L_{\rm TeV}\\
4\pi D_A^2\frac{dD_P}{dz'} 
\Theta\left[2-\Gamma_F(\Gamma_l,z')\right]
\tilde{\phi}_B(z',L_{\rm TeV},\Gamma_l)\,,
\label{eq:NB}
\end{multline}
where $F_{25,\rm min}\simeq5\times10^{-12}~{\rm erg~cm^{-2}~s^{-1}}$
is the flux limit of \Fermi~\citep[see Figures 15 \& 36 of ][]{2LAC}.
As with the \lnls~relation, the spectral cut on $\Gamma_F$ induces a
$\Gamma_l$-dependent redshift cutoff, limiting the potential
contributions from very bright objects at high-$z$.  This mimics the
luminosity upper limit we applied in Paper I, removing its
necessity\footnote{That such a limit exists, however, is strongly
  supported by the lack of a significant number HSPs in the 2LAC with
  $z>1$.  Note that unlike the hard gamma-ray blazars, the HSPs are
  defined by the location of the synchrotron peak, and are thus their
  definition is unaffected by the annihilation on the EBL suffered by
  the gamma rays.}.
In practice, we compare $d\log\N_B/dz$, both to avoid correlations in
the errors at subsequent redshifts and because the over-all
normalization has already been compared in the context of the
\lnls~relation.

Figure \ref{fig:dNdz} shows the $d\log\N_B/dz$ for \Fermi~hard
gamma-ray blazars with SNR $\ge 7$ in comparison to that inferred by
Equation (\ref{eq:NB}).  As in Paper I, the agreement is quite good,
though we miss what appears to be a small population of high-redshift
objects.  This may suggest an issue with our estimation of
$\Gamma_F(z,\Gamma_l)$, our distribution in $\Gamma_l$, and/or with the
source identification in the 2LAC at high $z$.  Alternatively, it may
suggest a possibly faster evolution for blazars in comparison with
quasars, as is the case for jet sources
\citep[i.e., radio-loud quasars, see, e.g.,][]{Singal+11,Singal+13}. In any case, it is 
clear that a rapidly evolving TeV blazar population is explicitly
consistent with the observed $d\log\N_B/dz$.  As with the \lnls~relation,
and unlike Paper I, there are no longer any free parameters.

\subsection{Isotropic Extragalactic Gamma-ray Background} \label{sec:EGRB}

In contrast to the \lnls~relation and redshift distribution of
\Fermi~blazars, the \Fermi~isotropic EGRB directly probes the population of
unresolved gamma-ray blazars at high-$z$.  Since a quasar-like
evolution of the blazar population has the most dramatic effect at
$z\sim1$--2, limits upon such an evolution have historically come from
modeling the EGRB.  

The \Fermi~EGRB spectrum is constructed following Section 5.3 of Paper
I, the only distinction being the subsequent average over $\Gamma_l$.
We first define a TeV-luminosity normalized intrinsic spectrum:
\begin{equation}
\hat{I}_E = \frac{E \,(dN/dE)}{
  \int_{0.1~{\rm TeV}}^{10~{\rm TeV}} dE \,E \, (dN/dE)
}\,,
\end{equation}
in terms of which, the EGRB spectrum is
\begin{multline}
E^2 \frac{d\bar{N}}{dE d\Omega}
=
\frac{E}{4\pi}
\int_0^2 d\Gamma_l
\int_{z_{\rm 2LAC}}^\infty dz
\int_{0}^{\infty}
d\log_{10} L_{\rm TeV}\\
4\pi D_A^2 \frac{dD_P}{dz} 
\frac{L_{\rm TeV}}{4\pi D_L^2}
\hat{I}_{E(1+z)}
e^{-\tau(E,z)}
\tilde{\phi}_B(z,L_{\rm TeV},\Gamma_l)\,,
\end{multline}
To exclude identifiable point sources we impose a lower redshift
cutoff set by when the peak of the luminosity
function, at isotropic equivalent luminosity of 
$\sim2\times10^{45}~{\rm erg~s^{-1}}$, passes the 2LAC, flux limit,
approximately at $z_{\rm 2LAC}\simeq0.291$.  This is slightly larger
than the $z_{\rm 1LAC}\simeq0.25$ chosen in Paper I owing to the
increased sensitivity limit of the 2LAC, which after two years should
have a flux limit roughly $\sqrt{2}$ lower and thus related by
$D_L^2(z_{\rm 2LAC}) = \sqrt{2} D_L^2(z_{\rm 1LAC})$. 

\begin{figure}
\begin{center}
\includegraphics[width=\columnwidth]{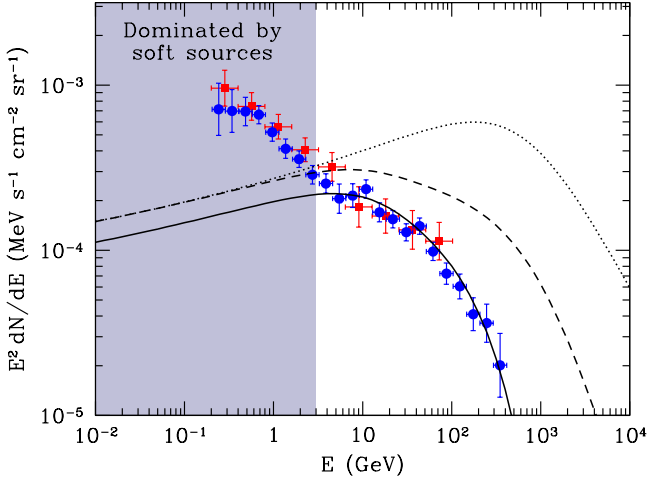}
\end{center}
\caption{\Fermi~isotropic EGRB anticipated by the
  hard gamma-ray blazars.  The dotted, dashed, and solid lines
  correspond to the unabsorbed spectrum, spectrum corrected for
  absorption on the EBL, and spectrum corrected for resolved point
  sources (assuming all hard gamma-ray blazars with $z\lesssim0.291$ are
  resolved, see text).  These are compared with the measured
  \Fermi~EGRB reported in \citet[][red squares]{Fermi_EGRB2010} and 
  \citet[][blue circles]{Fermi_EGRB2013}.  Note that below $\sim10$ GeV
  the EGRB is dominated by soft sources, specifically, the
  \Fermi~FSRQs.}\label{fig:EGRB} 
\end{figure}

Important modifications to the intrinsic hard gamma-ray blazar spectra
are the absorption above $\sim1$ TeV due to the extragalactic
background light (EBL) and the removal of identifiable point sources.
As mentioned earlier, it is in the treatment of the former that our
approach differs from other efforts to constrain the hard gamma-ray
blazar population: here we assume that this energy is primarily dumped
into the IGM as heat, instead of being reprocessed to lower energies
by the inverse Compton cascades.  Both the unabsorbed and point source
uncorrected spectra are shown in Figure \ref{fig:EGRB} for comparison.

\begin{figure}
\begin{center}
\includegraphics[width=\columnwidth]{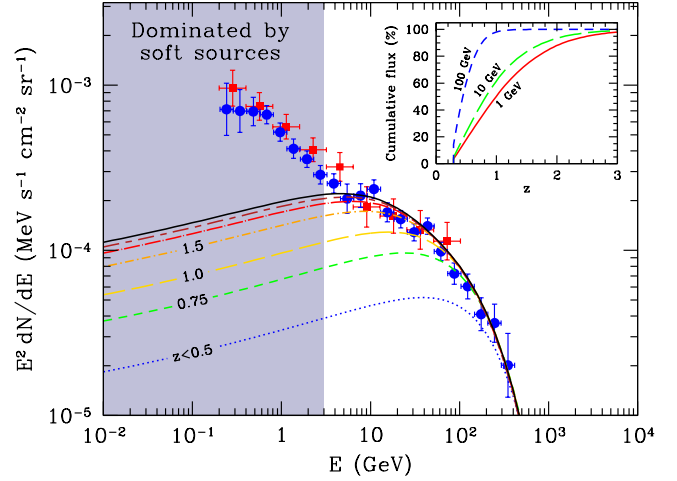}
\end{center}
\caption{Contribution to the \Fermi~EGRB from hard gamma-ray blazars
  below various redshifts.  Specifically, for objects with $z<0.5$
  (blue dot), 0.75 (green short dash), 1.0 (yellow long dash),
  1.5 (orange short dash dot), 2.0 (red long dash dot), 2.5 (dark red
  long dash-short dash), and all redshifts (black solid).  In all
  cases it was assumed that all sources with $z\lesssim0.291$ are
  resolved.
  These are compared with the measured
  \Fermi~EGRB reported in \citet[][red squares]{Fermi_EGRB2010} and 
  \citet[][blue circles]{Fermi_EGRB2013}.  Note that below $\sim10$ GeV
  the EGRB is dominated by soft sources, specifically, the
  \Fermi~FSRQs.
  The inset shows the cumulative flux fraction as a function of
  redshift for 1 GeV (red solid), 10 GeV (green long dash), and 100
  GeV (blue short dash).
}\label{fig:EGRB_z} 
\end{figure}

Combined with absorption and point source identification, the high
value of $\Gamma_h$ measured in bright TeV sources implies that the
EGRB must be substantially suppressed above a TeV.  
That is, the power-law behavior
implied by the \citet{Fermi_EGRB2010} measurement of the \Fermi~EGRB
cannot extend significantly beyond the 100 GeV upper limit for which
it was reported.  This is seen explicitly in Figure \ref{fig:EGRB},
where for the distribution of $\Gamma_l$ adopted in Section
\ref{sec:IS} the anticipated contribution to the EGRB peaks near 10
GeV followed by a rapid decline at larger photon energies.
This provides a remarkable agreement with the recent estimate of
\Fermi~EGRB spectrum by \citet{Fermi_EGRB2013}, shown by the blue
circles in in Figure \ref{fig:EGRB}.
Moreover, the spectrum of the EGRB above 6 GeV appears
to show a high energy bump upon a monotonically decreasing spectrum,
which we identify with the specific population of hard gamma-ray
blazars. 

Even in the presence of substantial absorption, the bulk of the EGRB
is produced at high redshifts, seen explicitly in Figure
\ref{fig:EGRB_z}.  At 10 GeV, roughly half of the observed EGRB is due
to objects with $z>1$.  The typical redshifts that contribute
are necessarily energy-dependent, with the higher energy EGRB arising
from more nearby sources.  Nevertheless, it is clear that even above a
few GeV the \Fermi~EGRB is probing the high-redshift blazar population.

The contribution to the EGRB spectrum from the hard gamma-ray blazars
below 10 GeV is nearly flat, and consequently dominates their
contribution to $\F_{25}$, consisting of roughly 10\%, and responsible
for the value obtained in Section \ref{sec:lNlS}.  However, this
number is quite uncertain, depending upon the behavior of the
extension of the gamma-ray blazar luminosity function to
$\Gamma_l>2$.

Below a few GeV the \Fermi~EGRB is dominated by soft gamma-ray
sources, the most important of which are the FSRQs 
\citep[see, e.g.,][]{Cavadini+2011,Stec-Vent:11}.  Their intrinsically
soft spectra combined with their typically larger luminosities (and
thus higher redshifts) confine their contribution to below $\sim3$
GeV.  Thus, our neglect of these sources is unlikely to significantly
change the \Fermi~EGRB above 10 GeV, where the hard gamma-ray blazars
successfully reproduce the observed background.

We note that the overall normalization of the TeV blazar
luminosity density is subject to an uncertain correction factor that
depends primarily on the incomplete census of the observed TeV
blazar population and enters linearly into the overall normalization
of the EGRB. We estimated this correction factor using the source
counts of hard Fermi blazars and confirm its value by comparison to
the redshift and cumulative flux distributions of these
objects. Nevertheless, there are remaining uncertainties associated
with the contribution of sources without measured redshifts and with
the extrapolation of that population of the \Fermi band to TeV
energies. If another plausible source population such as starburst
galaxies contributes a non-negligible, but subdominant, signal to the
EGRB, it could be
accommodated by a modest rescaling of the TeV luminosity density or 
slight modification to the hard gamma-ray blazar redshift evolution.
Despite this uncertainty,
the impressive match between the EGRB shape at energies above
$\sim3$~GeV strongly suggest that it is dominated by
a rapidly evolving hard gamma-ray blazar population.

\subsection{Anisotropy of the Extragalactic Gamma-ray Background} \label{sec:EGRB_aniso}

\begin{figure}
\begin{center}
\includegraphics[width=\columnwidth]{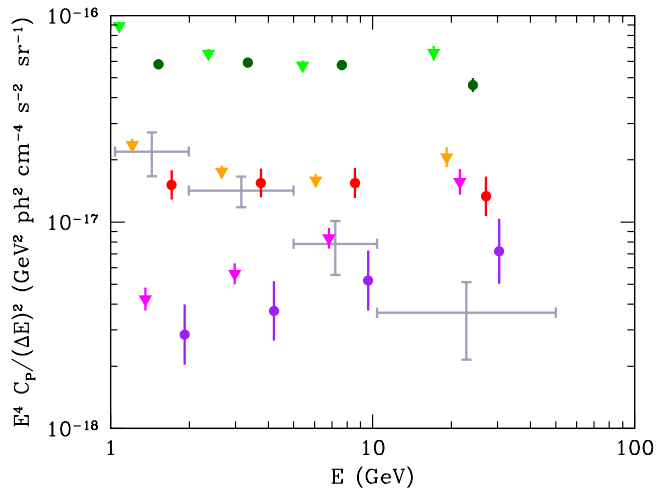}
\end{center}
\caption{
  Various estimates of the EGRB anisotropy spectrum compared with that
  anticipated by the hard gamma-ray blazar luminosity function in
  Equation (\ref{eq:eBLF}).  The estimates associated with the
  power-law extension of the 2FGL, the 2FGL alone, and hard sources
  ($\Gamma_F<2$) within the 2FGL, are shown by the green, orange, and
  magenta triangles (left to right).  The corresponding expectations
  from the hard gamma-ray blazars are shown by the dark green, red,
  and purple circles (left to right).  In all cases bars denote the
  1$\sigma$ cosmic variance uncertainty.  For reference, the grey
  bars show the energy bins employed and values reported in
  \citet{Fermi_aniso}, though see \citet{PaperVa} regarding a
  discussion of their normalization.  Points are horizontally offset
  within each bin for clarity.
}\label{fig:EA}
\end{figure}

In principle, the anisotropy of the EGRB limits the potential
contribution from discrete sources, providing a second direct
constraint on the fraction of the EGRB associated with blazars.  
The angular power in the EGRB on small angular
scales\footnote{Explicitly, multiples with $155\le\ell\le500$, above
  which the \Fermi PSF suppresses the angular power.} is observed to be
roughly constant, consistent with the expected Poisson noise due to an
unclustered population of point sources \citep{Fermi_aniso}.  The
magnitude of the angular power spectrum is energy dependent, yielding
an EGRB anisotropy spectrum, $C_P(E)$, shown by the grey bars in
Figure \ref{fig:EA}.

The origin of the constraint is straightforward to understand:
large numbers of blazars result in large Poisson fluctuations, and
therefore correspondingly large values of $C_P$.  Similarly, the
anisotropy spectrum's energy dependence is directly associated with the
underlying intrinsic spectra of the blazars: hard blazar spectra
produce hard anisotropy spectra.  The particular value of the EGRB
anisotropy depends, however, upon which sources are included.  Here we
follow \citet{Cuoco} and compute the anisotropy associated with
subsets of sources unresolved by the First Year \Fermi-LAT Source
Catalog (1FGL), corresponding roughly to a 1~GeV--100~GeV fluence
limit of $\F_{35}^{\rm 1FGL}=5\times10^{-10}\,{\rm ph~cm^{-2}~s^{-1}}$
(though see Appendix \ref{sec:FGLdeteff} for detailed estimates of the
fluence-dependent detection efficiency).

In terms of the TeV blazar luminosity function, the expected EGRB
anisotropy due to the hard gamma-ray blazars within an energy band
bounded by $E_m$ and $E_M$ is given by
\begin{multline}
C_{P,mM}
=
\int_0^\Gamma d\Gamma_l \int_0^\infty dz 
\int_{-\infty}^{\infty}
d\log_{10} L_{\rm TeV}\\
4\pi D_A^2 \frac{dD_P}{dz}
\F^2_{mM}
\tilde{\phi}_B(z,L_{\rm TeV},\Gamma_l)
w(\F_{35})
\,,
\label{eq:CPmM}
\end{multline}
where $\F_{mM}$ is the fluence in the specified energy band, specified
in Equation (\ref{eq:fF}), with the energy range explicitly
identified, and $w(\F_{35})$ is a weighting that describes the
detection efficiency for the sample under consideration (see below).

In practice, the normalization of the EGRB anisotropy spectrum reported
in \citet{Fermi_aniso} is inconsistent with the contributions arising
from sources already resolved in the 2 Year \Fermi LAT Source Catalog
(2FGL).  Detected sources in the 1FHL alone, without
correcting for the 1FHL detection efficiency, are sufficient to
account for the entirety of the reported EGRB anisotropy signal
above 10~GeV \citep{PaperVa}.  Thus, consistency with the reported EGRB
anisotropy would require a dramatic, and implausible, suppression in
the \lnls~relation, shown in Figure \ref{fig:lNlS_1FHL}, immediately
below the 1FHL detection threshold.
 
It is possible, however, to unambiguously compare the anticipated
hard gamma-ray blazar contribution to the EGRB anisotropy spectrum with
either that from the  2FGL sources alone (for which an unambiguous
estimate does exist) or simple extrapolations of the 2FGL source
population (providing a reasonable upper limit).  In the former, we
consider the contribution to the EGRB arising from blazars that lie
between the 2FGL and 1FGL detection thresholds.
In the latter we consider all sources below the 1FGL flux
limit, but compare the result to the EGRB anisotropy due to the
power-law extrapolation of the 2FGL fluence distribution described in
\citet{PaperVa}.

These comparisons are distinguished by the form of the weighting
function, $w(\F_{35})$, appearing in Equation (\ref{eq:CPmM}), which
in both cases may be constructed from the detection efficiencies of
the 1FGL and 2FGL, $\epsilon_{\rm 1FGL}(\F_{35})$ and $\epsilon_{\rm
  2FGL}(\F_{35})$, respectively (explicit expressions for these are
provided in Appendix \ref{sec:FGLdeteff}).  In the case of the
power-law extrapolation we need only to exclude
sources that are detected in the 1FGL, i.e.,
$w(\F_{35})=1-\epsilon_{\rm 1FGL}(\F_{35})$.  When comparing to the
2FGL contribution, we must also consider the probability that sources
are detected in the 2FGL, thus $w(\F_{35})=[1-\epsilon_{\rm
  1FGL}(\F_{35})]\epsilon_{\rm 2FGL}(\F_{35})$.

\begin{figure}
\begin{center}
\includegraphics[width=\columnwidth]{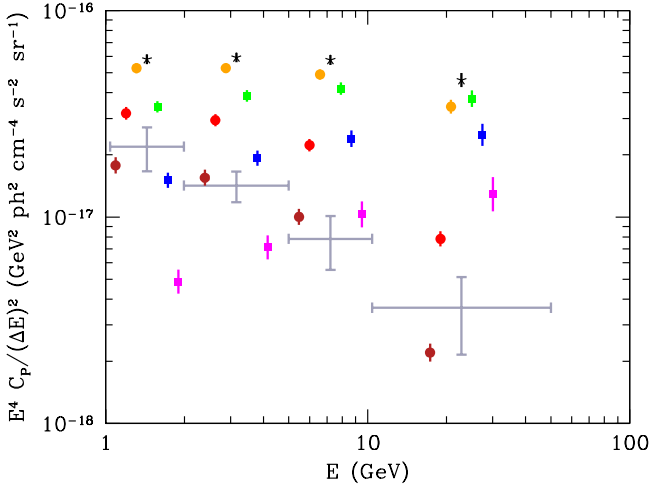}
\end{center}
\caption{
  Cumulative contributions of various redshifts and intrinsic photon
  spectral index ($\Gamma_l$) to the EGRB anisotropy spectra.
  Contributions from sources with $z>0.291$, 1, and
  1.5 are shown by the orange, red, and dark red circles, respectively
  (right to left within each energy bin). 
  Contributions from objects with $\Gamma_l<1.9$, 1.8, and 1.7, are
  shown by the green, blue, and magenta squares, respectively (left to
  right within each energy bin).
  For comparison, the black star shows the EGRB anisotropy spectrum
  associated with all sources.
  In all cases bars denote the 1$\sigma$ cosmic variance uncertainty.
  For reference, the grey bars show the energy bins employed and
  values reported in \citet{Fermi_aniso}, though see \citet{PaperVa}
  regarding a discussion of their normalization.  Points are horizontally
  offset within each bin for clarity.
}\label{fig:EAcum}
\end{figure}

The EGRB anisotropy implied by the hard gamma-ray blazar luminosity
function in Equation (\ref{eq:eBLF}) is significantly larger than the
values reported in \cite{Fermi_aniso} in all energy bands.  This is
unsurprising given that the reported values appear to be a substantial
underestimate of the EGRB anisotropy spectrum itself \citep{PaperVa}.
Nevertheless, it is consistent with (i.e., lies below) the values
anticipated from a smooth power-law extrapolation of the 2FGL, shown in 
Figure \ref{fig:EA}.  More importantly, a similar conclusion follows
from the comparison to the contribution to the EGRB anisotropy from the
2FGL sample alone.  This holds
for both the full 2FGL sample and the sub-sample of hard sources
(i.e., $\Gamma_F<2$).  Thus, despite being inconsistent with the reported
values in \citet{Fermi_aniso}, the hard gamma-ray blazar luminosity
function in Equation (\ref{eq:eBLF}) is able to reproduce the inferred
anisotropy signal associated with the currently known and smoothly
extrapolated point source samples, respectively.

The $C_{P,mM}$ are dominated by nearby sources, i.e., $z\lesssim1$.
This is clearly seen in Figure \ref{fig:EAcum}, in which the
contribution from sources with $z>1$ is below 33\% above 1.99~GeV, and
rapidly decreasing with redshift cut.  As a result, unlike the
isotropic EGRB component, the EGRB anisotropy is a probe of the
nearby blazar distribution (immediately below the 1FGL, and above the
2FGL, detection thresholds).
This is a direct result of the dominance of sources near
the detection threshold in the definition of the 
$C_{P,mM}$\footnote{The contribution of the hard gamma-ray blazars per
  logarithmic fluence interval to the EGRB anisotropy is 
  $d C_P/d\log\F = \F^3 d\N/d\F$.  Above and below the break of the
  \lnls~relation for the hard gamma-ray blazars we find
  $\N\propto\F^\alpha$ with $\alpha=-1.42$ and $-0.75$, respectively
  (see, e.g., Figures \ref{fig:lNlS} and \ref{fig:lNlS_1FHL}).
  Thus, generally the anisotropy is dominated by the population near
  the detection threshold, and is insensitive to the particulars of the
  population at substantially lower fluences.}. 
 Hence,
the agreement with the 2FGL contribution to the anisotropy is largely
anticipated by the success at reproducing the statistics of the
low-redshift hard gamma-ray blazar sample described in Sections
\ref{sec:lNlS} and \ref{sec:dNdz}.

The fractional contribution of intrinsically hard sources increases
with energy, though even in the highest energy bin (10.4~GeV--50~GeV)
sources with $\Gamma_l<\bar{\Gamma}_l$ contribute less than half of
the anisotropy signal (see the blue squares in Figure \ref{fig:EAcum}).  
Therefore, even a moderate restriction on $\Gamma_l$ produces a
substantial reduction in the anisotropy at all energies, implying that
our anisotropy estimates are sensitive to the high-$\Gamma_l$
extension of the gamma-ray blazar luminosity function.  Despite this
uncertainty, the hard gamma-ray blazars contribute substantially, if
not dominantly, to the anisotropy above roughly 3~GeV, consistent with
their contribution to the isotropic EGRB.  That is, it is possible to
simultaneously match both the isotropic and anisotropic components of
the EGRB with the single hard gamma-ray blazar population postulated here.

It is tempting to conclude that the absence of inverse Compton
cascades, which could be preempted by the presence of virulent plasma
beam instabilities, enables a notable consistency within the context
of the simplest model conceivable.  That is, the resolved source class of
hard gamma-ray blazars, which dominates the extragalactic high-energy
regime, also dominates the angular power as well as matches the
detailed shape and normalization of the isotropic EGRB intensity above
3~GeV.  However, as noted above, the current ambiguity in the
normalization of the reported EGRB anisotropy presently  precludes
such a statement in general.  The consistency obtained in Figure
\ref{fig:EA} is largely degenerate with the success in reproducing the
low-redshift gamma-ray blazar population.  Given the dominance of
low-redshift source contribution to the anisotropy, this is likely
to continue to be the case in the future.  As a result, even with the
dramatic evolution in the blazar population posited here, the EGRB
anisotropy will predominantly probe the low-redshift blazar
distribution generally.

\section{Conclusions} \label{sec:C}

In contrast to previous claims, a quasar-like evolution in the number density
of TeV blazars is fully consistent with the properties of the observed
\Fermi~population.  The chief uncertainties remain fundamentally
astrophysical: 1. How to relate the fluxes within the \Fermi-relevant
energy range and the intrinsic TeV luminosity, used to define the TeV
blazar luminosity function, and 2. The efficiency of the inverse
Compton cascades, if present at all.

A broken power-law model for the intrinsic TeV blazar spectrum, with a
generic break energy and high-energy photon spectral index of 1 TeV
and 3, respectively, is sufficient to reproduce many of the features
of the \Fermi~hard gamma-ray blazar population.  The TeV blazar
luminosity function was constructed for blazars that were observed at
TeV energies and hence is fundamentally 
limited to spectra which peak near $\sim1$ TeV, and thus we
necessarily impose an upper cutoff in the low-energy photon spectral
index of 2.  This cutoff is empirically supported by the observed
distribution of \Fermi~photon spectral indexes for the known TeV
blazars.

Modeling systematic biases is crucial to relating the intrinsic
blazar population and the \Fermi~blazar sample.  Of these, the most
important is the softening of the \Fermi-band spectra due to
absorption on the EBL, which causes a strong redshift-dependent
evolution in the observed photon spectral index from 1 GeV--100 GeV.
This, in turn, induces a significant sensitivity to the form of the
intrinsic spectrum below 1 TeV, generally, and in our case the
low-energy photon spectral index, specifically.  For this reason, to
obtain robust estimates of the anticipated \lnls~relation and redshift
distribution of nearby \Fermi~hard gamma-ray blazars, we found it
necessary to expand the definition of the TeV blazar luminosity
function to include the low-energy spectral index distribution.  This
is well approximated by a Gaussian peaked at a photon spectral index
of 1.78 and standard deviation 0.18.
Due to the redshift-dependent spectral softening, the \lnls~relation
and hard gamma-ray blazar redshift distribution both probe primarily
$z\lesssim1$.

The 2LAC \lnls~relation is well reproduced for 100~MeV--100~GeV fluxes
above $10^{-11}~{\rm erg~cm^{-2}~s^{-1}}$.  At smaller fluxes a
catalog-dependent flattening of the \lnls~relation suggests the
presence of an unidentified systematic effect similar to that described
by \citet{SPA12}. We predict the presence of a break in the hard
gamma-ray blazar \lnls~relation roughly at the current \Fermi~flux
limit, $5\times10^{-12}~{\rm erg~cm^{-2}~s^{-1}}$.  However, the
location of this break is determined primarily by objects near our
low-energy photon spectral index cutoff ($\Gamma_l=2$), and thus is
potentially sensitive to the unmodeled soft end of the TeV blazar
luminosity function.

Both the shape and magnitude of the 1FHL \lnls~relation for
10~GeV--500~GeV fluences above 
$4\times10^{-10}~{\rm ph~cm^{-2}~s^{-1}}$, presumably dominated by the
hard, gamma-ray bright objects of interest here, is excellently
reproduced, after correcting for the 1FHL detection efficiency.
Again, we predict a break in the \lnls~relation at fluxes near the
threshold, though the specific value is dependent upon the blazar
luminosity function near photon spectral index cutoff, and is
therefore somewhat uncertain.

Similarly, we are able to obtain a good fit to the \Fermi~2LAC
hard gamma-ray blazar redshift distribution.  In contrast to similar
calculations in Paper I, it is no longer necessary to specify an
arbitrary $\Gamma_l$ relationship between the inferred \Fermi-band
and TeV luminosities, or maximum intrinsic TeV luminosity,
substantially improving the robustness of the expected distribution.
In comparison to that from the 2LAC, our $d\log\N_B/dz$ falls
marginally faster, either due to our assumption of a fixed-flux cutoff
or suggesting an even more radical evolution of the TeV blazar
luminosity function at low redshift.

In contrast to the \lnls~relation and the hard gamma-ray blazar
redshift distribution, the \Fermi~EGRB directly probes the high-$z$
evolution of the TeV blazar luminosity function.  Below $\sim3$ GeV the
FSRQs, and other soft sources, dominate the EGRB.  However, above $\sim3$
GeV, where soft sources contribute negligibly, the expected
contribution from the hard gamma-ray blazars provide a remarkable fit
to the most recently reported \Fermi~EGRB.  
Of particular importance is the now observed strong suppression above
100 GeV; due to absorption on the EBL this is a robust prediction of
the TeV blazar luminosity function.  

Simultaneously, the hard gamma-ray blazars reproduce the observed
degree of anisotropy in the \Fermi~EGRB at energies where they
dominate the isotropic component.   This is possible since the
anisotropic and isotropic components of the EGRB are probing the hard
gamma-ray blazar population at different redshifts (being dominated by
nearby bright and distant dim objects, respectively), with the
disparity being precisely that anticipated by the rapidly evolving TeV
blazar luminosity function we have posited (note that this implies
this success may be largely degenerate with the ability to reproduce
the statistics of the hard gamma-ray blazars at low redshifts).  Thus,
above $\sim3$ GeV the \Fermi~EGRB may be fully explained within the
context of the single resolved source class of hard gamma-ray blazars.

\begin{deluxetable*}{cccccccc}
\tablecaption{Parameters of the Quasar Luminosity Function from \citet{Hopkins+07}\label{tab:QLF}}
\tablehead{
\multicolumn{2}{c}{Normalization}
&
\multicolumn{2}{c}{$\log_{10} L_*$}
&
\multicolumn{2}{c}{$\gamma_1$}
&
\multicolumn{2}{c}{$\gamma_2$}
}
\startdata
$\log_{10}\phi_*$ \tablenotemark{a} & $-4.825\pm0.060$
&
$\left(\log_{10} L_*\right)_0$ \tablenotemark{b} & $13.036\pm0.043$
&
$\gamma_{1,0}$ & $0.417\pm0.055$
&
$\gamma_{2,0}$ & $2.174\pm0.055$\\
&&
$k_{L,1}$ & $0.632\pm0.077$
&
$k_{\gamma_1}$ & $-0.623\pm0.132$
&
$k_{\gamma_2,1}$ & $1.460\pm0.096$\\
&&
$k_{L,2}$ & $-11.76\pm0.38$
&
&&
$k_{\gamma_2,2}$ & $-0.793\pm0.057$\\
&&
$k_{L,3}$ & $-14.25\pm0.80$
&&
&&\\
\enddata
\tablenotetext{a}{In units of co-moving $\Mpc^{-3}$}
\tablenotetext{b}{In units of $L_\odot\equiv3.9\times10^{33}\,\erg\,\s^{-1}$}
\end{deluxetable*}

The comparisons described above are based on an a priori model for the
TeV blazar population, with no adjustable parameters.  Thus, the
success of the TeV blazar luminosity function is non-trivial; these
are not ``fits'' in the normal sense.  However, critical to these is
the absence of the inverse Compton cascade emission that reprocesses
the flux above $\sim$TeV into the \Fermi-energy bands.  If this occurs,
the \Fermi~flux for a given TeV luminosity would increase
substantially, moving the \lnls~relations towards higher fluxes,
$d\log\N_B/dz$ towards higher $z$, the EGRB towards higher energy
fluxes, the EGRB anisotropy towards higher variances,
and thus in all cases badly violating the existing \Fermi~limits.  
Insofar as the evolution of TeV blazars may be expected to
qualitatively reflect the cosmological history of accretion on to
halos, this success may be seen as tentative support for the absence
of the inverse Compton cascades, and thus presumably circumstantial
evidence in favor of the existence of the only known alternative, beam
plasma instabilities.

\begin{appendix}

\section{An Explicit Expression for the Quasar Luminosity Function} \label{app:QLF}

In the interests of completeness, here we reproduce the co-moving
quasar luminosity function, $\QLF$ from \citet{Hopkins+07},
corresponding to the ``Full'' case in that paper, that we employ.  See
\citet{Hopkins+07} for how this $\QLF$ was obtained, and caveats
regarding its application.

The form of $\QLF$ is assumed to be a broken power law:
\begin{equation}
\QLF = \frac{\phi_*}{[L/L_*(z)]^{\gamma_1(z)}+[L/L_*(z)]^{\gamma_2(z)}}\,,
\end{equation}
where the location of the break ($L_*(z)$) and the power laws
($\gamma_1(z)$ and $\gamma_2(z)$) are functions of redshift.  These
are given by,
\begin{equation}
\begin{aligned}
\log_{10} L_*(z) &= \left(\log_{10} L_*\right)_0 + k_{L,1}\xi + k_{L,2}\xi^2 + k_{L,3}\xi^3\\
\gamma_1(z) &= \gamma_{1,0} 10^{k_{\gamma_1}\xi}\\
\gamma_2(z) &= 2 \gamma_{2,0} \left( 10^{k_{\gamma_2,1}\xi} + 10^{k_{\gamma_2,2}\xi} \right)^{-1}\\
\end{aligned}
\end{equation}
where
\begin{equation}
\xi \equiv \log_{10}\left(\frac{1+z}{3}\right)\,.
\end{equation}
The values of the relevant parameters are given in Table
\ref{tab:QLF}.  Finally, $\tilde{\phi}_Q$, defined in terms of
physical volume, is related in the usual way:
\begin{equation}
\tilde{\phi}_Q(z,L)
=
(1+z)^3\QLF\,.
\end{equation}

\section{Explicit Flux Definitions} \label{app:fdefs}
We employ three definitions of ``flux'' here: the fluences
from 100 MeV--100 GeV ($\F_{25}$) and 1 GeV--100 GeV ($\F_{35}$)
and the flux from 100 MeV--100 GeV ($F_{25}$).  These are related to
the reported $\F_{35}$ via Equations (\ref{eq:fFdef}) and
(\ref{eq:Fdef}).  Explicitly, setting $f_F$ with by $\F_{35}$, 
\begin{equation}
f_F = \frac{\Gamma_F-1}{1 - 100^{1-\Gamma_F}} \F_{35}~{\rm GeV^{\Gamma_F-1} s^{-1}}\,,
\end{equation}
the corresponding value for $\F_{25}$ is
\begin{equation}
\F_{25} 
= 
\frac{0.1^{1-\Gamma_F}-100^{1-\Gamma_F}}{1-100^{1-\Gamma_F}} \F_{35}\,,
\label{eq:f25}
\end{equation}
where we have assumed $\Gamma_F\ne1$.
Similarly, $F_{25}$ is given by
\begin{equation}
F_{25} 
= 
\left\{
\begin{aligned}
&\frac{\Gamma_F-1}{\Gamma_F-2} \frac{0.1^{2-\Gamma_F}-100^{2-\Gamma_F}}{1-100^{1-\Gamma_F}} \F_{35}
~{\rm GeV}
&&
\Gamma_F\ne2\\
&\frac{\Gamma_F-1}{1-100^{1-\Gamma_F}} 
\log(10^3)
\F_{35}
~{\rm GeV}
&&
\Gamma_F=2\,,
\end{aligned}
\right.
\label{eq:F2535}
\end{equation}
where the additional factor of a GeV sets the scale of the energy flux.

\section{Detection Efficiencies of High Latitude Gamma-ray Point Source Samples}
Here we summarize the detection efficiencies associated with various
high-latitude point source samples employed in the text.

\subsection{1FHL} \label{sec:1FHLdeteff}

\begin{figure}
\begin{center}
\includegraphics[width=0.8\columnwidth]{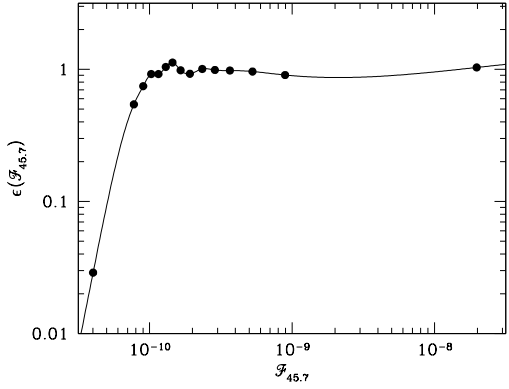}
\end{center}
\caption{1FHL Detection Efficiency for high-latitude sources
  ($|b|>15^\circ$), taken from Figure 30 of \citet{1FHL}.}\label{fig:deteff}
\end{figure}

In the construction of the 1FHL \lnls~relation we make an attempt to
account for the detection efficiency using the values shown in Figure
30 of \citet{1FHL}.  Specifically, we set
\begin{equation}
\N(S) = \sum_j \frac{1}{\epsilon(\F_{45.7})} \Theta(\F_{45.7}-S)
\end{equation}
where $\epsilon(\F_{45.7})$ is the spline-interpolated detection
efficiency shown in Figure \ref{fig:deteff}.

\subsection{1FGL and 2FGL} \label{sec:FGLdeteff}

\begin{figure}
\begin{center}
\includegraphics[width=0.8\columnwidth]{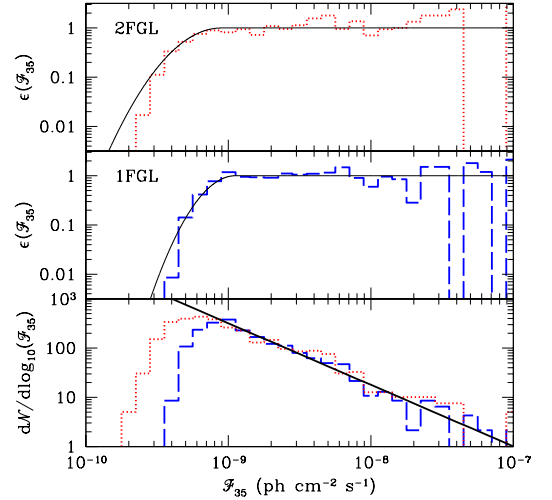}
\end{center}
\caption{1FGL and 2FGL Detection Efficiency for high-latitude sources
  ($|b|>15^\circ$).  Bottom: $\F_{35}$ distributions of the 1FGL (blue
  dashed) and 2FGL (red dotted) in comparison to a power-law
  extrapolation of the latter (black solid).  Middle and top:
  Estimates of the detection efficiencies of the 1FGL and 2FGL
  assuming the power-law extrapolation of the 2FGL approximates the
  true source flux distribution.}\label{fig:1FGL2FGLdeteff}
\end{figure}

Unlike the 1FHL, the 1FGL and 2FGL point source detection efficiency
are not present in the literature.  However, these are necessary to
reconstruct the anticipated EGRB anisotropy spectra associated with
populations either masked by, or due to, these populations.  Here we
approximately reconstruct these detection efficiencies, following the
procedure employed in \citet{PaperVa} (to which we direct the reader
for a more complete discussion).

Figure \ref{fig:1FGL2FGLdeteff} shows the distribution of all sources
with $|b|>15^\circ$ in the 1FGL and 2FGL catalogs in $\F_{35}$.  This
flux measure was chosen since it is both reported in the 1FGL and 2FGL
catalogs and is apparently uncorrelated with the photon spectral
index.
Above $\F_{35}\simeq10^{-9}~{\rm ph~cm^{-2}~s^{-1}}$ the two
populations are both consistent with a single power law,
$\propto\F_{35}^{-1.25}$.  At lower fluences the number of 1FGL
rapidly decreases.  The extension of the 2FGL to even lower fluences,
where it to exhibits a rapid decline, implies that these are
associated with the detection efficiency of the respective catalogs
and not with some intrinsic feature of the underlying source
population.

Assuming that the high-fluence power law provides an approximation of
the true source population, the ratio of the observed fluence
distribution to the power law provides an estimate of the desired
detection efficiency, $\epsilon(\F_{35})$.  These are shown in the top
two panels of Figure \ref{fig:1FGL2FGLdeteff}.  We approximate the
detection efficiencies by a
\begin{equation}
\epsilon(\F_{35})
=
\begin{cases}
10^{-m [\log_{10}(\F_{35}/\F_{\rm max})]^2} & \F_{35}<\F_{\rm max}\\
1 & \text{otherwise}\,,
\end{cases}
\end{equation}
where for the 1FGL we have $m^{\rm 1FGL}=7$ and 
$\F_{\rm max}^{\rm 1FGL}=1.12\times10^{-9}~{\rm ph~cm^{-2}~s^{-1}}$,
and for the 2FGL we have  $m^{\rm 2FGL}=4$ and 
$\F_{\rm max}^{\rm 2FGL}=0.89\times10^{-9}~{\rm ph~cm^{-2}~s^{-1}}$.
These fits, shown in Figure \ref{fig:1FGL2FGLdeteff}, are most
accurate in the immediate vicinity of the detection threshold, the
region that dominates the contribution to the EGRB anisotropy
measurements.

In these no attempt to correct for Eddington bias
\citep{Eddi:1913,Eddi:1940} has
been made, despite being evident in the 1FHL and 1FGL (resulting in
$\epsilon>1$ near the fluence threshold, corresponding to
lower-fluence sources being detected at higher fluences).  Doing so
would reduce the inferred EGRB anisotropies.

\end{appendix}

\acknowledgments The authors thank Markus Ackermann and the
\Fermi~collaboration for providing the preliminary \Fermi~EGRB
spectrum and Volker Springel for careful reading of the
manuscript.  A.E.B.~receives financial support from the Perimeter
Institute for Theoretical Physics and the Natural Sciences and
Engineering Research Council of Canada through a Discovery Grant.
Research at Perimeter Institute is supported by the Government of
Canada through Industry Canada and by the Province of Ontario through
the Ministry of Research and Innovation.  C.P.~gratefully acknowledges
financial support of the Klaus Tschira Foundation. E.P. acknowledges
support by the DFG through Transregio 33.  P.C. gratefully
acknowledges support from the UWM Research Growth Initiative, from
\Fermi Cycle 5 through NASA grant NNX12AP24G, from the NASA ATP
program through NASA grant NNX13AH43G, and NSF grant AST-1255469.

\bibliography{bigmh.bib,bigmh_orig.bib}

\begin{thebibliography}{33}
\expandafter\ifx\csname natexlab\endcsname\relax\def\natexlab#1{#1}\fi

\bibitem[{{Abdo} {et~al.}(2010{\natexlab{a}}){Abdo}, {Ackermann}, {Ajello},
  {Atwood}, {Baldini}, {Ballet}, {Barbiellini}, {Bastieri}, {Baughman},
  {Bechtol}, {Bellazzini}, {Berenji}, {Blandford}, {Bloom}, {Bonamente},
  {Borgland}, {Bregeon}, {Brez}, {Brigida}, {Bruel}, {Burnett}, {Buson},
  {Caliandro}, {Cameron}, {Caraveo}, {Casandjian}, {Cavazzuti}, {Cecchi}, {{\c
  C}elik}, {Charles}, {Chekhtman}, {Cheung}, {Chiang}, {Ciprini}, {Claus},
  {Cohen-Tanugi}, {Cominsky}, {Conrad}, {Cutini}, {Dermer}, {de Angelis}, {de
  Palma}, {Digel}, {di Bernardo}, {E Silva}, {Drell}, {Drlica-Wagner},
  {Dubois}, {Dumora}, {Farnier}, {Favuzzi}, {Fegan}, {Focke}, {Fortin},
  {Frailis}, {Fukazawa}, {Funk}, {Fusco}, {Gaggero}, {Gargano}, {Gasparrini},
  {Gehrels}, {Germani}, {Giebels}, {Giglietto}, {Giommi}, {Giordano},
  {Glanzman}, {Godfrey}, {Grenier}, {Grondin}, {Grove}, {Guillemot}, {Guiriec},
  {Gustafsson}, {Hanabata}, {Harding}, {Hayashida}, {Hughes}, {Itoh},
  {Jackson}, {J{\'o}hannesson}, {Johnson}, {Johnson}, {Johnson}, {Johnson},
  {Kamae}, {Katagiri}, {Kataoka}, {Kawai}, {Kerr}, {Kn{\"o}dlseder}, {Kocian},
  {Kuehn}, {Kuss}, {Lande}, {Latronico}, {Lemoine-Goumard}, {Longo}, {Loparco},
  {Lott}, {Lovellette}, {Lubrano}, {Madejski}, {Makeev}, {Mazziotta},
  {McConville}, {McEnery}, {Meurer}, {Michelson}, {Mitthumsiri}, {Mizuno},
  {Moiseev}, {Monte}, {Monzani}, {Morselli}, {Moskalenko}, {Murgia}, {Nolan},
  {Norris}, {Nuss}, {Ohsugi}, {Omodei}, {Orlando}, {Ormes}, {Paneque},
  {Panetta}, {Parent}, {Pelassa}, {Pepe}, {Pesce-Rollins}, {Piron}, {Porter},
  {Rain{\`o}}, {Rando}, {Razzano}, {Reimer}, {Reimer}, {Reposeur}, {Ritz},
  {Rochester}, {Rodriguez}, {Roth}, {Ryde}, {Sadrozinski}, {Sanchez}, {Sander},
  {Parkinson}, {Scargle}, {Sellerholm}, {Sgr{\`o}}, {Shaw}, {Siskind}, {Smith},
  {Smith}, {Spandre}, {Spinelli}, {Starck}, {Strickman}, {Strong}, {Suson},
  {Tajima}, {Takahashi}, {Takahashi}, {Tanaka}, {Thayer}, {Thayer}, {Thompson},
  {Tibaldo}, {Torres}, {Tosti}, {Tramacere}, {Uchiyama}, {Usher}, {Vasileiou},
  {Vilchez}, {Vitale}, {Waite}, {Wang}, {Winer}, {Wood}, {Ylinen}, {Ziegler},
  \& {Fermi-LAT Collaboration}}]{Fermi_EGRB2010}
{Abdo}, A.~A., {et~al.} 2010{\natexlab{a}}, \prl, 104, 101101

\bibitem[{{Abdo} {et~al.}(2010{\natexlab{b}}){Abdo}, {Ackermann}, {Ajello},
  {Allafort}, {Antolini}, {Atwood}, {Axelsson}, {Baldini}, {Ballet},
  {Barbiellini}, {Bastieri}, {Baughman}, {Bechtol}, {Bellazzini}, {Berenji},
  {Blandford}, {Bloom}, {Bogart}, {Bonamente}, {Borgland}, {Bouvier},
  {Bregeon}, {Brez}, {Brigida}, {Bruel}, {Buehler}, {Burnett}, {Buson},
  {Caliandro}, {Cameron}, {Cannon}, {Caraveo}, {Carrigan}, {Casandjian},
  {Cavazzuti}, {Cecchi}, {{\c C}elik}, {Celotti}, {Charles}, {Chekhtman},
  {Chen}, {Cheung}, {Chiang}, {Ciprini}, {Claus}, {Cohen-Tanugi}, {Conrad},
  {Costamante}, {Cotter}, {Cutini}, {D'Elia}, {Dermer}, {de Angelis}, {de
  Palma}, {De Rosa}, {Digel}, {Silva}, {Drell}, {Dubois}, {Dumora}, {Escande},
  {Farnier}, {Favuzzi}, {Fegan}, {Ferrara}, {Focke}, {Fortin}, {Frailis},
  {Fukazawa}, {Funk}, {Fusco}, {Gargano}, {Gasparrini}, {Gehrels}, {Germani},
  {Giebels}, {Giglietto}, {Giommi}, {Giordano}, {Giroletti}, {Glanzman},
  {Godfrey}, {Grandi}, {Grenier}, {Grondin}, {Grove}, {Guiriec}, {Hadasch},
  {Harding}, {Hayashida}, {Hays}, {Healey}, {Hill}, {Horan}, {Hughes},
  {Iafrate}, {Itoh}, {J{\'o}hannesson}, {Johnson}, {Johnson}, {Johnson},
  {Johnson}, {Kamae}, {Katagiri}, {Kataoka}, {Kawai}, {Kerr}, {Kn{\"o}dlseder},
  {Kuss}, {Lande}, {Latronico}, {Lavalley}, {Lemoine-Goumard}, {Llena Garde},
  {Longo}, {Loparco}, {Lott}, {Lovellette}, {Lubrano}, {Madejski}, {Makeev},
  {Malaguti}, {Massaro}, {Mazziotta}, {McConville}, {McEnery}, {McGlynn},
  {Michelson}, {Mitthumsiri}, {Mizuno}, {Moiseev}, {Monte}, {Monzani},
  {Morselli}, {Moskalenko}, {Murgia}, {Nolan}, {Norris}, {Nuss}, {Ohno},
  {Ohsugi}, {Omodei}, {Orlando}, {Ormes}, {Ozaki}, {Paneque}, {Panetta},
  {Parent}, {Pelassa}, {Pepe}, {Pesce-Rollins}, {Piranomonte}, {Piron},
  {Porter}, {Rain{\`o}}, {Rando}, {Razzano}, {Reimer}, {Reimer}, {Reposeur},
  {Ripken}, {Ritz}, {Rodriguez}, {Romani}, {Roth}, {Ryde}, {Sadrozinski},
  {Sanchez}, {Sander}, {Saz Parkinson}, {Scargle}, {Sgr{\`o}}, {Shaw},
  {Siskind}, {Smith}, {Spandre}, {Spinelli}, {Starck}, {Stawarz}, {Strickman},
  {Suson}, {Tajima}, {Takahashi}, {Takahashi}, {Tanaka}, {Taylor}, {Thayer},
  {Thayer}, {Thompson}, {Tibaldo}, {Torres}, {Tosti}, {Tramacere}, {Ubertini},
  {Uchiyama}, {Usher}, {Vasileiou}, {Vilchez}, {Villata}, {Vitale}, {Waite},
  {Wallace}, {Wang}, {Winer}, {Wood}, {Yang}, {Ylinen}, \&
  {Ziegler}}]{Fermi_AGNCatalogue2010}
---. 2010{\natexlab{b}}, \apj, 715, 429

\bibitem[{{Abdo} {et~al.}(2010{\natexlab{c}}){Abdo}, {Ackermann}, {Agudo},
  {Ajello}, {Aller}, {Aller}, {Angelakis}, {Arkharov}, {Axelsson}, {Bach},
  {et~al.}}]{Fermi_SED2010}
---. 2010{\natexlab{c}}, \apj, 716, 30

\bibitem[{{Ackermann} {et~al.}(2011){Ackermann}, {Ajello}, {Allafort},
  {Antolini}, {Atwood}, {Axelsson}, {Baldini}, {Ballet}, {Barbiellini},
  {Bastieri}, {Bechtol}, {Bellazzini}, {Berenji}, {Blandford}, {Bloom},
  {Bonamente}, {Borgland}, {Bottacini}, {Bouvier}, {Bregeon}, {Brigida},
  {Bruel}, {Buehler}, {Burnett}, {Buson}, {Caliandro}, {Cameron}, {Caraveo},
  {Casandjian}, {Cavazzuti}, {Cecchi}, {Charles}, {Cheung}, {Chiang},
  {Ciprini}, {Claus}, {Cohen-Tanugi}, {Conrad}, {Costamante}, {Cutini}, {de
  Angelis}, {de Palma}, {Dermer}, {Digel}, {Silva}, {Drell}, {Dubois},
  {Escande}, {Favuzzi}, {Fegan}, {Ferrara}, {Finke}, {Focke}, {Fortin},
  {Frailis}, {Fukazawa}, {Funk}, {Fusco}, {Gargano}, {Gasparrini}, {Gehrels},
  {Germani}, {Giebels}, {Giglietto}, {Giommi}, {Giordano}, {Giroletti},
  {Glanzman}, {Godfrey}, {Grenier}, {Grove}, {Guiriec}, {Gustafsson},
  {Hadasch}, {Hayashida}, {Hays}, {Healey}, {Horan}, {Hou}, {Hughes},
  {Iafrate}, {J{\'o}hannesson}, {Johnson}, {Johnson}, {Kamae}, {Katagiri},
  {Kataoka}, {Kn{\"o}dlseder}, {Kuss}, {Lande}, {Larsson}, {Latronico},
  {Longo}, {Loparco}, {Lott}, {Lovellette}, {Lubrano}, {Madejski}, {Mazziotta},
  {McConville}, {McEnery}, {Michelson}, {Mitthumsiri}, {Mizuno}, {Moiseev},
  {Monte}, {Monzani}, {Moretti}, {Morselli}, {Moskalenko}, {Murgia},
  {Nakamori}, {Naumann-Godo}, {Nolan}, {Norris}, {Nuss}, {Ohno}, {Ohsugi},
  {Okumura}, {Omodei}, {Orienti}, {Orlando}, {Ormes}, {Ozaki}, {Paneque},
  {Parent}, {Pesce-Rollins}, {Pierbattista}, {Piranomonte}, {Piron}, {Pivato},
  {Porter}, {Rain{\`o}}, {Rando}, {Razzano}, {Razzaque}, {Reimer}, {Reimer},
  {Ritz}, {Rochester}, {Romani}, {Roth}, {Sanchez}, {Sbarra}, {Scargle},
  {Schalk}, {Sgr{\`o}}, {Shaw}, {Siskind}, {Spandre}, {Spinelli}, {Strong},
  {Suson}, {Tajima}, {Takahashi}, {Takahashi}, {Tanaka}, {Thayer}, {Thayer},
  {Thompson}, {Tibaldo}, {Tinivella}, {Torres}, {Tosti}, {Troja}, {Uchiyama},
  {Vandenbroucke}, {Vasileiou}, {Vianello}, {Vitale}, {Waite}, {Wallace},
  {Wang}, {Winer}, {Wood}, {Wood}, \& {Zimmer}}]{2LAC}
{Ackermann}, M., {et~al.} 2011, \apj, 743, 171

\bibitem[{{Ackermann} {et~al.}(2012{\natexlab{a}}){Ackermann}, {Ajello},
  {Albert}, {Baldini}, {Ballet}, {Barbiellini}, {Bastieri}, {Bechtol},
  {Bellazzini}, {Bloom}, {Bonamente}, {Borgland}, {Brandt}, {Bregeon},
  {Brigida}, {Bruel}, {Buehler}, {Buson}, {Caliandro}, {Cameron}, {Caraveo},
  {Cecchi}, {Charles}, {Chekhtman}, {Chiang}, {Ciprini}, {Claus},
  {Cohen-Tanugi}, {Conrad}, {Cuoco}, {Cutini}, {D'Ammando}, {de Palma},
  {Dermer}, {Digel}, {do Couto e Silva}, {Drell}, {Drlica-Wagner}, {Dubois},
  {Favuzzi}, {Fegan}, {Ferrara}, {Fortin}, {Fukazawa}, {Fusco}, {Gargano},
  {Gasparrini}, {Germani}, {Giglietto}, {Giroletti}, {Glanzman}, {Godfrey},
  {Gomez-Vargas}, {Gr{\'e}goire}, {Grenier}, {Grove}, {Guiriec}, {Gustafsson},
  {Hadasch}, {Hayashida}, {Hayashi}, {Hou}, {Hughes}, {J{\'o}hannesson},
  {Johnson}, {Kamae}, {Kn{\"o}dlseder}, {Kuss}, {Lande}, {Latronico},
  {Lemoine-Goumard}, {Linden}, {Lionetto}, {Llena Garde}, {Longo}, {Loparco},
  {Lovellette}, {Lubrano}, {Mazziotta}, {McEnery}, {Mitthumsiri}, {Mizuno},
  {Monte}, {Monzani}, {Morselli}, {Moskalenko}, {Murgia}, {Naumann-Godo},
  {Norris}, {Nuss}, {Ohsugi}, {Okumura}, {Orienti}, {Orlando}, {Ormes},
  {Paneque}, {Panetta}, {Parent}, {Pavlidou}, {Pesce-Rollins}, {Pierbattista},
  {Piron}, {Pivato}, {Rain{\`o}}, {Rando}, {Reimer}, {Reimer}, {Roth},
  {Sbarra}, {Schmitt}, {Sgr{\`o}}, {Siegal-Gaskins}, {Siskind}, {Spandre},
  {Spinelli}, {Strong}, {Suson}, {Takahashi}, {Tanaka}, {Thayer}, {Tibaldo},
  {Tinivella}, {Torres}, {Tosti}, {Troja}, {Usher}, {Vandenbroucke},
  {Vasileiou}, {Vianello}, {Vitale}, {Waite}, {Winer}, {Wood}, {Wood}, {Yang},
  {Zimmer}, \& {Komatsu}}]{Fermi_aniso}
---. 2012{\natexlab{a}}, \prd, 85, 083007

\bibitem[{{Ackermann} {et~al.}(2012{\natexlab{b}})}]{Fermi_EGRB2013}
---. 2012{\natexlab{b}}, 4th Fermi Symposium

\bibitem[{{Ackermann} {et~al.}(2012{\natexlab{c}}){Ackermann}, {Ajello},
  {Allafort}, {Schady}, {Baldini}, {Ballet}, {Barbiellini}, {Bastieri},
  {Bellazzini}, {Blandford}, {Bloom}, {Borgland}, {Bottacini}, {Bouvier},
  {Bregeon}, {Brigida}, {Bruel}, {Buehler}, {Buson}, {Caliandro}, {Cameron},
  {Caraveo}, {Cavazzuti}, {Cecchi}, {Charles}, {Chaves}, {Chekhtman}, {Cheung},
  {Chiang}, {Chiaro}, {Ciprini}, {Claus}, {Cohen-Tanugi}, {Conrad}, {Cutini},
  {D'Ammando}, {de Palma}, {Dermer}, {Digel}, {do Couto e Silva},
  {Dom{\'{\i}}nguez}, {Drell}, {Drlica-Wagner}, {Favuzzi}, {Fegan}, {Focke},
  {Franckowiak}, {Fukazawa}, {Funk}, {Fusco}, {Gargano}, {Gasparrini},
  {Gehrels}, {Germani}, {Giglietto}, {Giordano}, {Giroletti}, {Glanzman},
  {Godfrey}, {Grenier}, {Grove}, {Guiriec}, {Gustafsson}, {Hadasch},
  {Hayashida}, {Hays}, {Jackson}, {Jogler}, {Kataoka}, {Kn{\"o}dlseder},
  {Kuss}, {Lande}, {Larsson}, {Latronico}, {Longo}, {Loparco}, {Lovellette},
  {Lubrano}, {Mazziotta}, {McEnery}, {Mehault}, {Michelson}, {Mizuno}, {Monte},
  {Monzani}, {Morselli}, {Moskalenko}, {Murgia}, {Tramacere}, {Nuss},
  {Greiner}, {Ohno}, {Ohsugi}, {Omodei}, {Orienti}, {Orlando}, {Ormes},
  {Paneque}, {Perkins}, {Pesce-Rollins}, {Piron}, {Pivato}, {Porter},
  {Rain{\`o}}, {Rando}, {Razzano}, {Razzaque}, {Reimer}, {Reimer}, {Reyes},
  {Ritz}, {Rau}, {Romoli}, {Roth}, {S{\'a}nchez-Conde}, {Sanchez}, {Scargle},
  {Sgr{\`o}}, {Siskind}, {Spandre}, {Spinelli}, {Stawarz}, {Suson},
  {Takahashi}, {Tanaka}, {Thayer}, {Thompson}, {Tibaldo}, {Tinivella},
  {Torres}, {Tosti}, {Troja}, {Usher}, {Vandenbroucke}, {Vasileiou},
  {Vianello}, {Vitale}, {Waite}, {Winer}, {Wood}, \& {Wood}}]{Fermi_EBL2012}
---. 2012{\natexlab{c}}, Science, 338, 1190

\bibitem[{{Ackermann} {et~al.}(2013){Ackermann}, {Ajello}, {Allafort}, ,
  {Atwood}, {Baldini}, {Ballet}, {Barbiellini}, {Bastieri}, {Bechtol}, \&
  et~al.}]{1FHL}
---. 2013, ArXiv e-prints

\bibitem[{{Broderick} {et~al.}(2012){Broderick}, {Chang}, \&
  {Pfrommer}}]{PaperI}
{Broderick}, A.~E., {Chang}, P., \& {Pfrommer}, C. 2012, \apj, 752, 22

\bibitem[{{Broderick} {et~al.}(2013)}]{PaperVa}
{Broderick}, A.~E., {et~al.} 2013, {\em submitted to} \apj

\bibitem[{{Cavadini} {et~al.}(2011){Cavadini}, {Salvaterra}, \&
  {Haardt}}]{Cavadini+2011}
{Cavadini}, M., {Salvaterra}, R., \& {Haardt}, F. 2011, arXiv:1105.4613

\bibitem[{{Chang} {et~al.}(2012){Chang}, {Broderick}, \& {Pfrommer}}]{PaperII}
{Chang}, P., {Broderick}, A.~E., \& {Pfrommer}, C. 2012, \apj, 752, 23

\bibitem[{{Cuoco} {et~al.}(2012){Cuoco}, {Komatsu}, \&
  {Siegal-Gaskins}}]{Cuoco}
{Cuoco}, A., {Komatsu}, E., \& {Siegal-Gaskins}, J.~M. 2012, \prd, 86, 063004

\bibitem[{{Eddington}(1913)}]{Eddi:1913}
{Eddington}, A.~S. 1913, \mnras, 73, 359

\bibitem[{{Eddington}(1940)}]{Eddi:1940}
{Eddington}, Sir, A.~S. 1940, \mnras, 100, 354

\bibitem[{{Ghisellini}(2011)}]{Ghis:11}
{Ghisellini}, G. 2011, arXiv: 1104.0006

\bibitem[{{Gould} \& {Schr{\'e}der}(1967)}]{Goul-Schr:67}
{Gould}, R.~J., \& {Schr{\'e}der}, G.~P. 1967, Physical Review, 155, 1408

\bibitem[{{Hopkins} {et~al.}(2007){Hopkins}, {Richards}, \&
  {Hernquist}}]{Hopkins+07}
{Hopkins}, P.~F., {Richards}, G.~T., \& {Hernquist}, L. 2007, \apj, 654, 731

\bibitem[{{Inoue} \& {Totani}(2009)}]{Inou-Tota:09}
{Inoue}, Y., \& {Totani}, T. 2009, \apj, 702, 523

\bibitem[{{Kneiske} {et~al.}(2004){Kneiske}, {Bretz}, {Mannheim}, \&
  {Hartmann}}]{Knei_etal:04}
{Kneiske}, T.~M., {Bretz}, T., {Mannheim}, K., \& {Hartmann}, D.~H. 2004, \aap,
  413, 807

\bibitem[{{Kneiske} \& {Mannheim}(2008)}]{Knei-Mann:08}
{Kneiske}, T.~M., \& {Mannheim}, K. 2008, \aap, 479, 41

\bibitem[{{Miniati} \& {Elyiv}(2012)}]{Mini-Elyv:12}
{Miniati}, F., \& {Elyiv}, A. 2012, ArXiv e-prints

\bibitem[{{Narumoto} \& {Totani}(2006)}]{Naru-Tota:06}
{Narumoto}, T., \& {Totani}, T. 2006, \apj, 643, 81

\bibitem[{{Neronov} \& {Semikoz}(2009)}]{Nero-Semi:09}
{Neronov}, A., \& {Semikoz}, D.~V. 2009, \prd, 80, 123012

\bibitem[{{Pfrommer} {et~al.}(2012){Pfrommer}, {Chang}, \&
  {Broderick}}]{PaperIII}
{Pfrommer}, C., {Chang}, P., \& {Broderick}, A.~E. 2012, \apj, 752, 24

\bibitem[{{Puchwein} {et~al.}(2012){Puchwein}, {Pfrommer}, {Springel},
  {Broderick}, \& {Chang}}]{PaperIV}
{Puchwein}, E., {Pfrommer}, C., {Springel}, V., {Broderick}, A.~E., \& {Chang},
  P. 2012, \mnras, 423, 149

\bibitem[{{Salamon} \& {Stecker}(1998)}]{Sala-Stec:98}
{Salamon}, M.~H., \& {Stecker}, F.~W. 1998, \apj, 493, 547

\bibitem[{{Schlickeiser} {et~al.}(2012){Schlickeiser}, {Ibscher}, \&
  {Supsar}}]{Schl_etal:12}
{Schlickeiser}, R., {Ibscher}, D., \& {Supsar}, M. 2012, \apj, 758, 102

\bibitem[{{Singal} {et~al.}(2012){Singal}, {Petrosian}, \& {Ajello}}]{SPA12}
{Singal}, J., {Petrosian}, V., \& {Ajello}, M. 2012, \apj, 753, 45

\bibitem[{{Singal} {et~al.}(2011){Singal}, {Petrosian}, {Lawrence}, \&
  {Stawarz}}]{Singal+11}
{Singal}, J., {Petrosian}, V., {Lawrence}, A., \& {Stawarz}, {\L}. 2011, \apj,
  743, 104

\bibitem[{{Singal} {et~al.}(2013){Singal}, {Petrosian}, {Stawarz}, \&
  {Lawrence}}]{Singal+13}
{Singal}, J., {Petrosian}, V., {Stawarz}, {\L}., \& {Lawrence}, A. 2013, \apj,
  764, 43

\bibitem[{{Stecker} \& {Venters}(2011)}]{Stec-Vent:11}
{Stecker}, F., \& {Venters}, T.~M. 2011, \apj, 736, 40

\bibitem[{{Venters}(2010)}]{Vent:10}
{Venters}, T.~M. 2010, \apj, 710, 1530

\end{thebibliography}
\bibliographystyle{apj}

\end{document}